\documentclass[a4paper,12pt]{article}
\DeclareUnicodeCharacter{2032}{\ensuremath{^{\prime}}} 
\usepackage{pstricks}
\usepackage{color}
\usepackage{amssymb,amsmath,bm,bbm}
\usepackage{epsf,epsfig}
\usepackage{afterpage}
\usepackage{longtable}
\usepackage{cite}
\usepackage{latexsym,mathrsfs,dsfont}
\usepackage{graphicx}
\usepackage{url}
\usepackage{paralist}
\usepackage{bbold}
\usepackage{appendix}
\usepackage{hyperref}
\usepackage{booktabs}
\usepackage{float}
\usepackage{comment}

\usepackage{tikz}
\usepackage{tikz-feynman}

\usepackage{soul}

\definecolor{dgreen}{rgb}{0.0, 0.5, 0.0}

\begin{document}

\hfill {\tt }

\def\thefootnote{\fnsymbol{footnote}}

\begin{center}
\Large\bf\boldmath
\vspace*{1.cm} 
An insight into the rare $Z\rightarrow b \bar{b}\gamma$ at the HL-LHC
\end{center}
\vspace{0.6cm}

\begin{center}
 T. Thallapalli$^{1}$\footnote{tejaswini.thallapalli@physics.iitd.ac.in},
 A.~Alpana$^{2}$\footnote{alpana.sirohi@students.iiserpune.ac.in},
 A.M. ~Iyer$^{1}$\footnote{iyerabhishek@physics.iitd.ac.in}, 
S.~Sharma$^{2}$\footnote{seema@iiserpune.ac.in}\\

\vspace{0.6cm}
{\sl $^1$Department of Physics, Indian Institute of Technology Delhi,
Hauz Khas, New Delhi-110016, India}\\[0.4cm]
{\sl $^2$Indian Institute of Science Education and Research, Pune-411008, India}
\\[0.4cm]

\end{center}
\renewcommand{\thefootnote}{\arabic{footnote}}
\setcounter{footnote}{0}

\vspace{1.cm}
\begin{abstract}
\noindent
Studies at the $Z$-pole have played an important role in developing our understanding of the Standard Model (SM). Continuing the explorations in this regime, 
we consider the possibility of the production of two $b$-quarks  and a photon in proton-proton collisions at the HL-LHC. While such a final state is possible in the SM by means of the process $Z\rightarrow b\bar b$ decay with a radiated photon, the focus is on extracting its possible origins due to beyond  Standard Model (BSM) physics. The signal topology can be broadly identified as $Z\rightarrow \Phi\gamma\rightarrow b\bar b\gamma$, where $
\Phi$ can either be a spin-0 or spin-2 state with a mass less than that of the $Z$ boson.
The analysis is characterised by two relatively low $p_T$ jets that are required to be $b$-tagged jets and an isolated photon. The study provides a quantitative framework for their identification and highlights the potential challenges associated with this final state.
A range of machine learning architectures is employed to demonstrate the stability and reliability of the discriminating variables.  This study highlights the importance of the low-$p_T$ objects in searches at the HL-LHC, hence the need to pay special attention to their identification and efficiencies.
\end{abstract}

\newpage

\section{Introduction}
The discovery of weak neutral currents has served as the cornerstone for the establishment of the $SU(2)_L\times U(1)_Y$ model as the theory of nature at the electroweak scale \cite{GargamelleNeutrino:1974khg}. Since then, 
 the study of the physics of the $Z$-pole at Large Electron Positron (LEP) Collider \cite{de_Boer_2015} and Stanford Linear Collider (SLC) \cite{SLCPhinney:2016uad} has played an important role in further shaping our understanding of this model that has come to be known as the Standard Model (SM) of particle physics. The high-precision measurements of different observables and their consistency with the SM prediction is a testimony to the success of the theory at the electroweak scale\cite{Altarelli:1997et}. 
 While these measurements are consistent with the total decay width $\Gamma_Z$, the possibility of additional rare decay channels remains, potentially becoming detectable with greater precision. For example, compared to the $\sim 10^6$ $Z$ bosons at the LEP and SLC, the proposed future circular electron-positron collider (FCCee) \cite{FCC:2018byv,FCC:2018evy} and Circular Electron Positron Collider (CEPC) \cite{CEPCStudyGroup:2023quu} anticipates $\sim 10^{12}$ $Z$ boson events at the end of their respective runs at the Tera-Z phase. This opens the door to exploring rare decays with branching fractions below $10^{-7}$ \cite{Cacciapaglia:2021agf}. 
 
 Of particular interest is the decay of the $Z$ boson into a three-body final state. In the realm of the Standard Model, searches for $Z\rightarrow\phi\gamma$ and $Z\rightarrow \eta\gamma$ exist where $\phi,\eta$ are the QCD mesons that further decay into lighter mesons (like Kaons) \cite{ATLAS:2016zln,ATLAS:2017gko}.   
  A three-body final state can also be admitted by means of a topology of the form $Z\rightarrow \Phi+\text{V}$, where $\Phi$ corresponds to a new physics state and `V' is a Standard Model particle. Depending on the underlying model, $\Phi$ can then further decay into a pair of SM particles, eventually resulting in a three-body final state. 
In principle,  $\Phi$ can admit both a fermionic and a bosonic nature. With the requirement that the state $V$ is a photon, the possibilities for $\Phi$ narrows down to either a spin-0 (scalar and pseudo-scalar) and spin-2 (Kaluza-Klein graviton). Additional spin-0 particles can emerge as scalars/pseudo-scalars in models with extended Higgs sectors\cite{Branco:2011iw,Ellwanger:2009dp}, composite models \cite{Cacciapaglia2014_composite} and extra-dimensional models\cite{Kribs:2006mq,Cheng:2010pt}.
  For the spin-0 possibility of $\Phi$, the most stringent upper bound of $\sim 10^{-4}$ applies to the di-lepton decay mode $\Phi \rightarrow \ell^+ \ell^-$ \cite{ParticleDataGroup:2024cfk}. The spin-0 scenario with different decay modes of $\Phi$ has been the subject of considerable investigation at proposed lepton colliders such as FCC-ee \cite{Cacciapaglia:2021agf} and CEPC. \cite{Xiao_2016}.
spin-2 states are characteristic of extra-dimensional frameworks, arising as Kaluza–Klein excitations of higher-dimensional gravity theories \cite{Arkani-Hamed:1998jmv, Randall:1999ee}.
  The gravitational decay of the $Z$ boson has been considered in the context of large extra-dimensional models \cite{Gabrielli_2002,Allanach:2007ea}. 

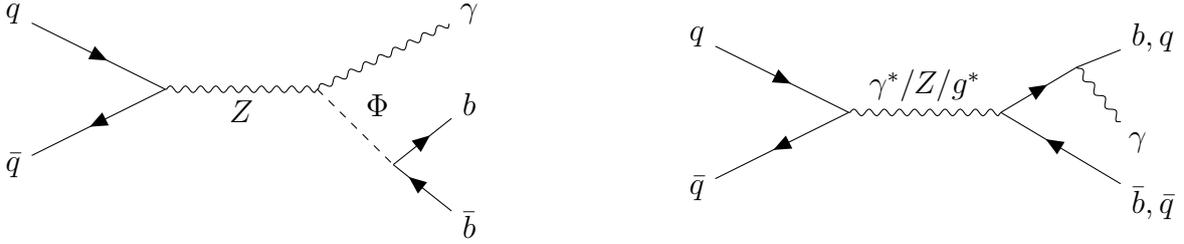
\begin{figure}[htb!]
\centering

\begin{minipage}{0.45\textwidth}
\centering
\begin{tikzpicture}
  \begin{feynman}
    
    \vertex (i1) at (-2,1) {\(q\)};
    \vertex (i2) at (-2,-1) {\(\bar q\)};
    \vertex (v1) at (0,0);
    \vertex (v2) at (2,0);
    \vertex (o1) at (4,1) {\(\gamma\)};
    \vertex (v3) at (3,-1);
    \vertex (o2) at (4,-0.2) {\(b\)};
    \vertex (o3) at (4,-1.8) {\(\bar b\)};

    \diagram*{
      (i1) -- [fermion] (v1) -- [fermion] (i2),
      (v1) -- [boson, edge label'=\(Z\)] (v2),
      (v2) -- [boson] (o1),
      (v2) -- [scalar, edge label=\(\Phi\)] (v3),
      (v3) -- [fermion] (o2),
      (v3) -- [anti fermion] (o3),
    };
  \end{feynman}
\end{tikzpicture}
\end{minipage}
\hfill
\begin{minipage}{0.45\textwidth}
\centering
\begin{tikzpicture}
  \begin{feynman}
    \vertex (i1) at (-2,1) {\(q\)};
    \vertex (i2) at (-2,-1) {\(\bar q\)};
    \vertex (v1) at (0,0);
    \vertex (v2) at (2,0);
    \vertex (bmid) at (3,0.6); 
    \vertex (o1) at (4,1.0) {\(b,q\)};
    \vertex (o2) at (4,-1.2) {\(\bar b,\bar{q}\)};
    \vertex (g1) at (3.8,-0.4) {\(\gamma\)};

    \diagram*{
      (i1) -- [fermion] (v1) -- [fermion] (i2),
      (v1) -- [boson, edge label=\(\gamma^*/Z/g^*\)] (v2),
      (v2) -- [fermion] (bmid) -- (o1),
      (v2) -- [anti fermion] (o2),
      (bmid) -- [photon] (g1),
    };
  \end{feynman}
\end{tikzpicture}
\end{minipage}

 \caption{Signal and background Feynman diagrams.}
\label{fig:feynman}
\end{figure}

The LEP and the future colliders represent two temporal extremes that exist in the context of the investigation of physics at the $Z$-pole. The high luminosity phase of the Large Hadron Collider (HL-LHC), with an estimated $\sim 10^9$ $ Z$-boson events, represents an ideal transition phase from the legacy of LEP to future experiments. This work will focus on the potential sensitivity of the LHC to the rare $Z\rightarrow b\bar b\gamma$ mode through $Z\rightarrow\Phi\gamma$ and $\Phi \rightarrow b\bar b$ channel as shown on the left side of Fig. \ref{fig:feynman}. The essence of the analysis lies in the fact that the results apply to both spin-0 and spin-2 hypotheses for $\Phi$. The mass of $\Phi$ ($m_\Phi$) is treated as a free parameter, assumed to be smaller than the $Z$-boson mass. The event is characterized by the presence of two $b$-tagged jets and an isolated photon. The analysis elucidates the challenges posed by the hadronic environment, particularly the identification of low transverse momentum ($p_T$) $b$-tagged jets and potentially soft photons. By employing strategies ranging from Boosted Decision Trees (BDTs) to advanced Graph Neural Networks (GNNs), we establish the strongest bounds on the branching fraction as a function of the photon transverse momentum ($p_T^\gamma$), thereby paving the way for future collider analyses.

The paper is organized as follows: In section \ref{sec:theoreticalorigins}, we outline the relevant theoretical frameworks and demonstrate the broad scope of the analysis that follows. 
 Section \ref{sec:variables} identifies the appropriate collider strategy by quantifying the efficiencies. It also specifies the discriminating features of interest to be used for the different ML architectures in Section \ref{sec:models}: A) Boosted Decision Trees (BDT) and B) Graph Neural Networks (GNN) with different message passing methods. The results are presented in Section \ref{sec:results} and followed by the conclusion.

\section{Theoretical Origins}
\label{sec:theoreticalorigins}
The simplest realization of a two-$b$-quark plus photon final state from $Z$-boson decay in the SM arises from the process $q\bar q \rightarrow Z \rightarrow b\bar b$, with the photon radiated from either the initial- or final-state particles, as illustrated in the right plot of Fig.~\ref{fig:feynman}. An additional background source is from the matrix element $\mathcal{M}(q\bar q \rightarrow q\bar q)$, where the $b$-jets may originate either from gluon splitting or from misidentified jets. 
However, attributing  a beyond Standard Model (BSM) origin due to the  
Z-boson decay to this final state requires the introduction of an additional coupling between the 
Z-boson, a light BSM particle ($\Phi$), and a photon ($\gamma$), as shown in the left plot of Fig.\ref{fig:feynman}. The light state would then eventually decay into a pair of $b$-quarks. It is interesting to note that there exist two broad possibilities for the spin of the light state $\Phi$: spin-0 corresponding to a (pseudo-)scalar or spin-2 corresponding to a graviton. The former is characteristic of scenarios like extended Higgs sectors\cite{Branco:2011iw}, NMSSM\cite{Ellwanger_2010}, composite models \cite{Cacciapaglia2014_composite,Cacciapaglia:2020kgq}, \textit{etc}. The latter, on the other hand, can find its origins in flat and warped extra-dimensional models \cite{Kribs:2006mq,Gherghetta:2010cj}. Regardless of the spin assignment of the BSM state $\Phi$, the most general effective Lagrangian at a new-physics scale $\Lambda$ parameterizing the $Z$–$\Phi$–$\gamma$ vertex takes the form
\begin{equation}
    \mathcal{L}\supset\frac{1}{\Lambda}  \left(\Phi_sZ_{\mu\nu}F^{\mu\nu}+ \Phi_{ps}\epsilon_{\mu\nu\alpha\beta} Z^{\alpha\beta}F^{\mu\nu}\right)+\frac{1}{\Lambda}h_{\mu\nu}T^{\mu\nu}_G
\label{eq:lagrangian}
\end{equation}
The first two terms correspond to the scalar ($s$) and pseudo-scalar ($ps$) nature of $\Phi$, respectively. For the spin-2 case, the vertex can be extracted from the final term where $T^{\mu\nu}_G$ is the stress-energy tensor for the gauge fields and $h_{\mu\nu}$ is the gravitational perturbation about the flat metric. 
The coupling to fermions can be written as
\begin{equation}
     \mathcal{L}\supset \Phi_s\bar ff +\Phi_{ps}\bar f\gamma_5f+\frac{1}{\Lambda}h_{\mu\nu}T^{\mu\nu}_F
\end{equation}
where $T^{\mu\nu}_F$ is the stress-energy tensor for the fermionic fields. Several scenarios may be characterized by the presence of both spin-0 and spin-2 states, {i.e}, more than one particle of the type $\Phi$ that may have a $Z$-$\Phi$-$\gamma$ vertex.
In such cases, we assume that only one of them satisfies the condition $m_\Phi<m_Z$. Thus, an analysis focusing on the decay $Z\rightarrow \Phi\gamma$ and $\Phi\rightarrow b\bar b$ translates into a search for the rare $Z\rightarrow b\bar b\gamma$ decay mode. For the analysis, BR of $\Phi\rightarrow b\bar b=1$ is assumed.

\begin{figure}[htb!]
    \centering
    \includegraphics[width=0.48\linewidth]{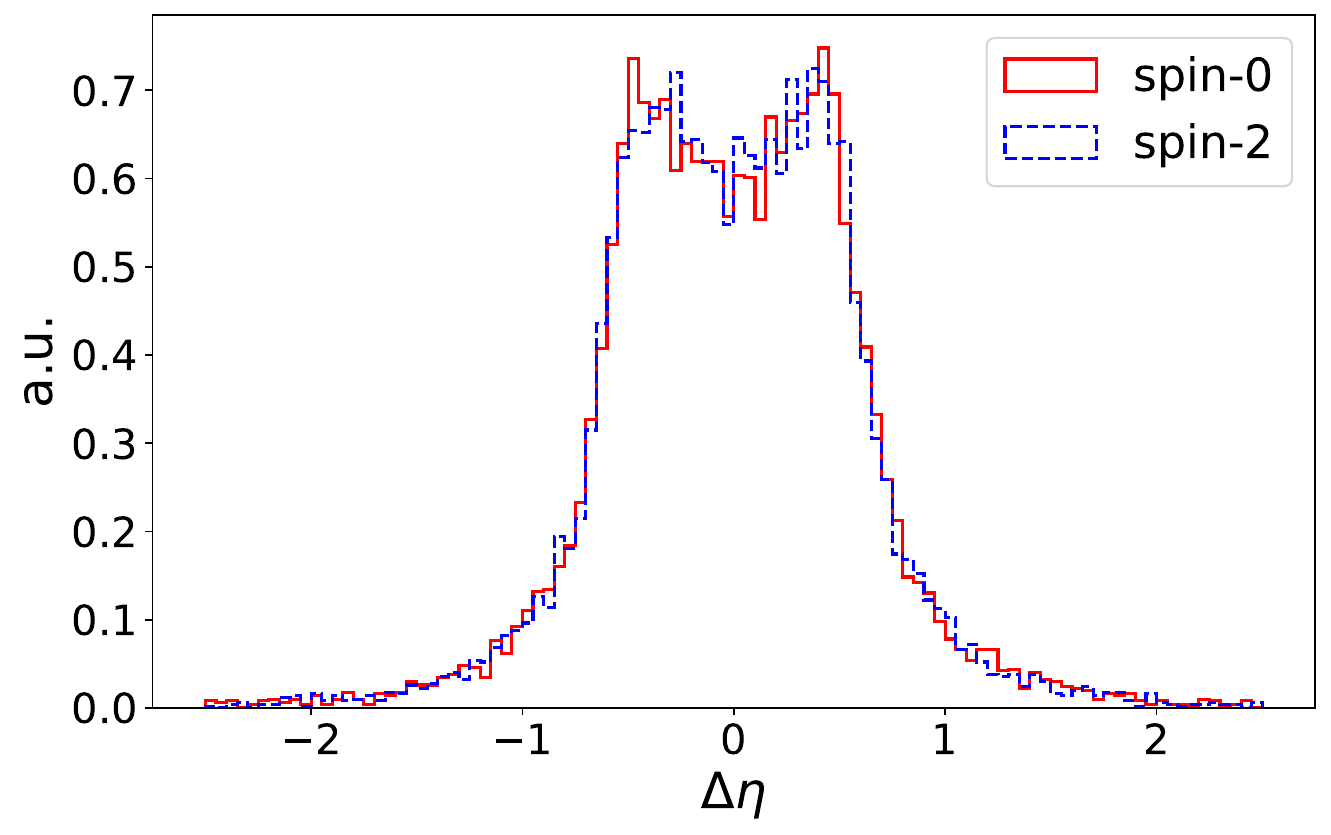}
    \includegraphics[width=0.48\linewidth]{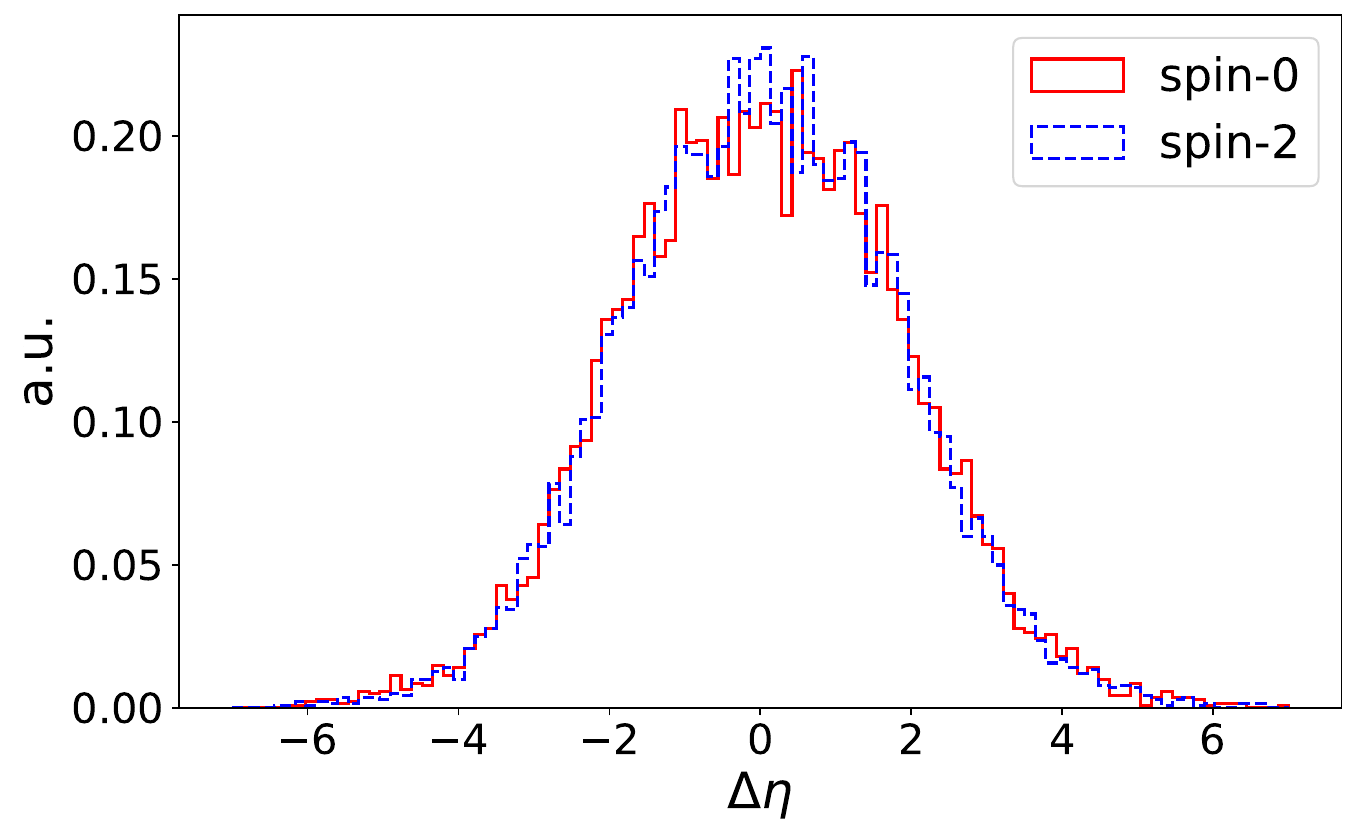}
    
    \caption{$\Delta\eta$ distribution between the two $b$-quarks  (left), $\Phi$ and $\gamma$ (right) for both the  spin-0 and spin-2 hypothesis.}
    \label{fig:deltaeta}
\end{figure}

  The development of an analysis that is broadly independent of the nature of $\Phi$ must ensure similar acceptance efficiencies when the standard signal selection criteria are imposed. The event will be characterized by two $b$-jets and an isolated photon. The detector efficiencies for a given final state are closely linked to the pseudo-rapidity  $\eta$ distribution of the particle produced from the matrix element.   
 We consider the $\Delta \eta$ distributions, as illustrated in Fig. \ref{fig:deltaeta}, between the two $b$ quarks (left) and between the $\Phi$ and the photon (right).
This can be extracted by generating the model file with {\tt FEYNRULES} \cite{Alloul:2013bka}, simulating the events using {\tt MADGRAPH} \cite{Alwall_2011}, and subsequently employing the particle PID numbers to obtain the distribution.  
 Since the scalar and pseudo-scalar cases exhibit identical patterns, we adopt the pseudo-scalar hypothesis as the representative for spin-0, without loss of generality. Estimating the correlation between the final states in both the primary and the secondary vertex would capture potential differences in the final efficiencies.
 The statistical agreement of the distributions between the spin-0 and spin-2 hypotheses supports the broad applicability of a unified analysis.

Thus, the only key difference from the background with a radiated photon is the presence of a light intermediate state. Its identification by means of its mass reconstruction and other unique kinematic features would play an important role in differentiating it from the background. The following sections will develop a collider strategy that would quantify the sensitivity of the LHC for such a signal.

\section{ Discriminants and Efficiencies}
\label{sec:variables}
The independence of event generation from the underlying signal hypothesis significantly simplifies the collider strategy. The analysis is performed under the spin-0 pseudo-scalar hypothesis; however, the results are, without loss of generality, directly applicable to alternative scenarios as well. The simulated matrix element is passed through {\tt PYTHIA} \cite{Bierlich:2022pfr} for showering and hadronization, following which the CMS card of {\tt DELPHES} \cite{2014} is used to get the detector output. The events are characterized by the presence of jets that are reconstructed using the AK4\cite{Cacciari_2008} clustering algorithm using {\tt FASTJET} \cite{Cacciari:2011ma} and isolated photons.  We impose the requirement that at least two of the jets are $b$-tagged, incorporating the relevant $p_T$-dependent tagging efficiencies for their identification. It is important to note that it is highly unlikely that there are more than two $b$-tagged jets in either a signal or background event. In order to maximize signal acceptance efficiency, the two jets identified as $b$-jets need not necessarily be the two leading jets in the event.
As the interest is a three-body final state, with two distinct objects: $\gamma$ and $b$-jets, it is instructive to understand the individual roles played by each in influencing the overall event selection efficiencies.

\begin{figure}[htb!]
    \centering
    \includegraphics[width=0.48\linewidth]{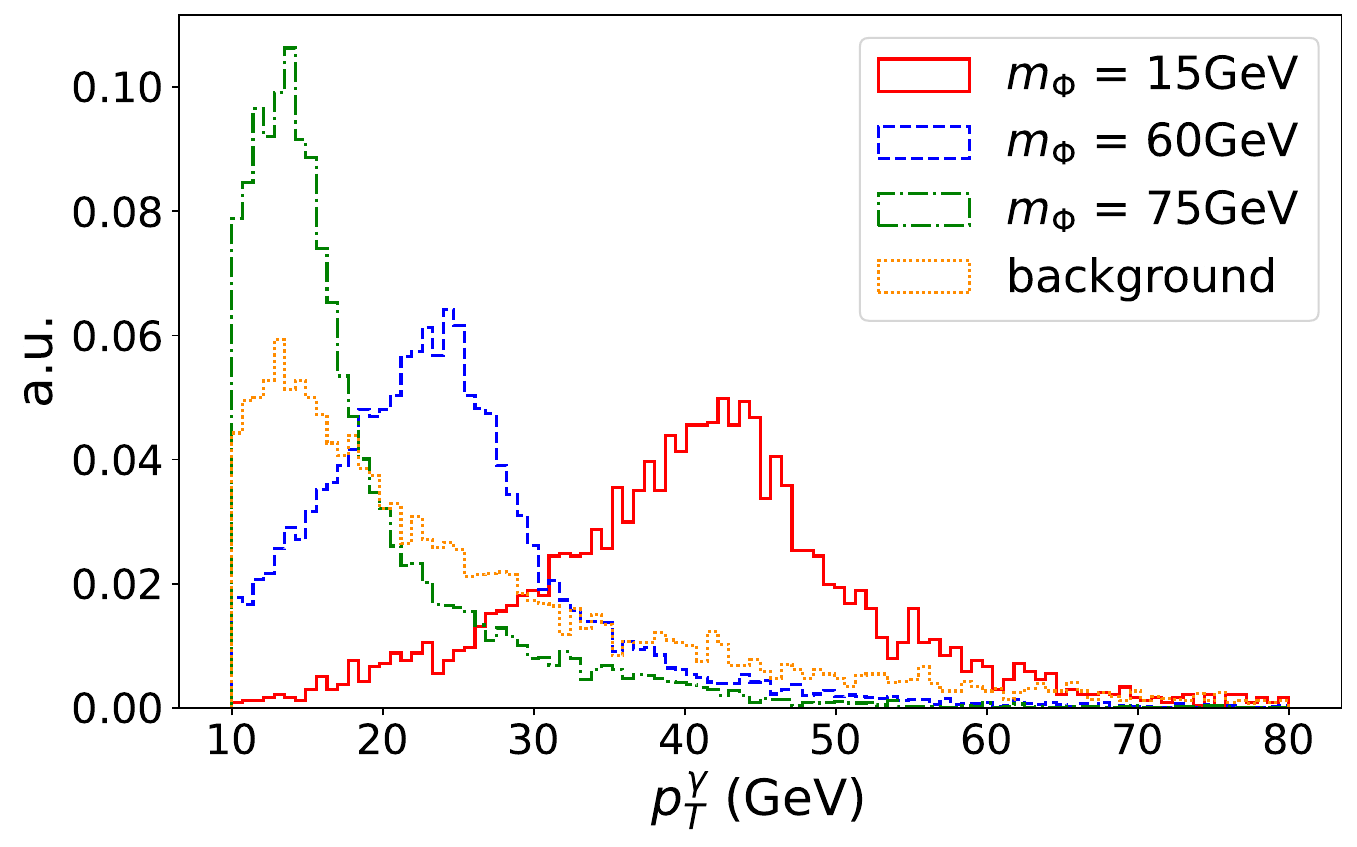} 
    \includegraphics[width=0.5\linewidth]{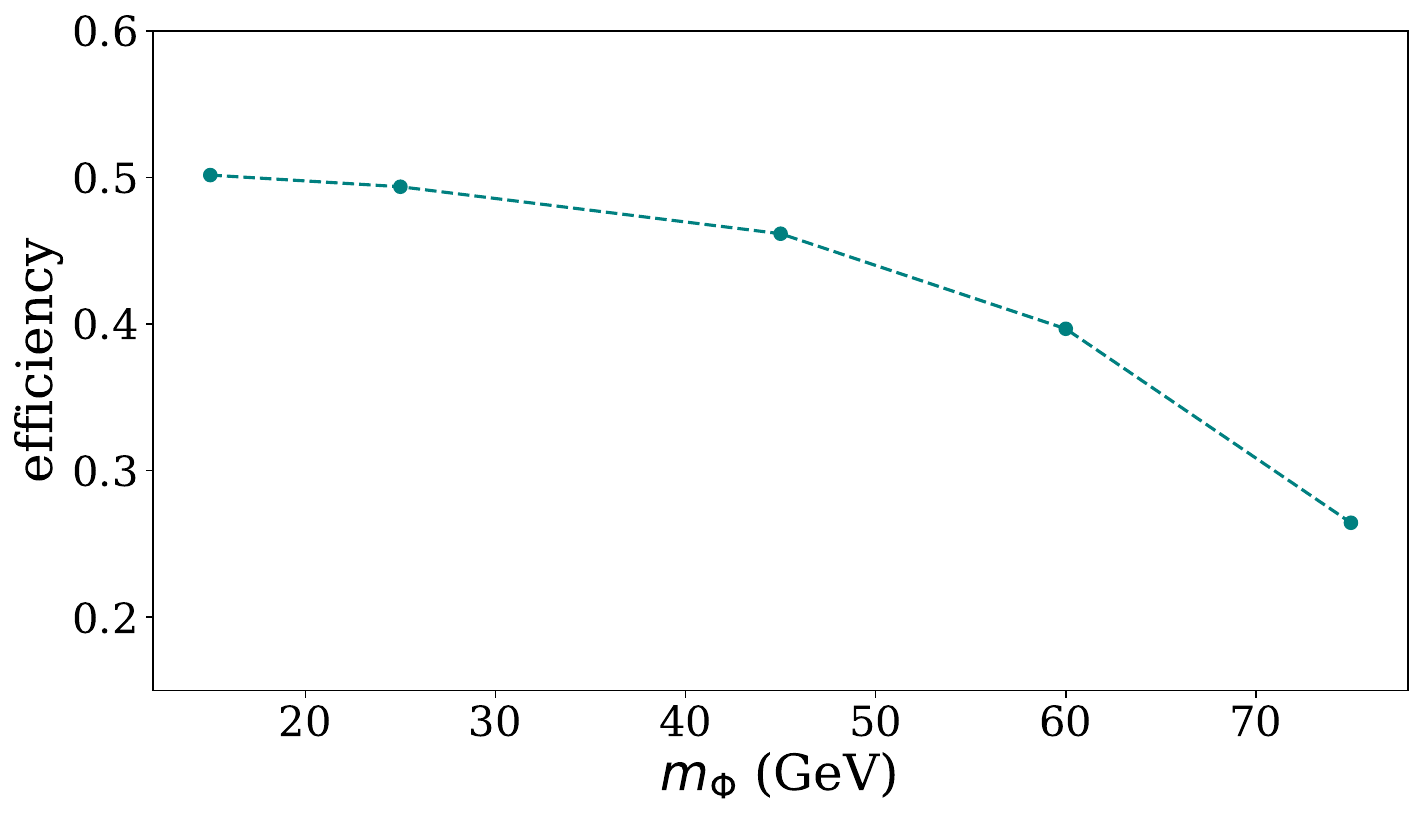}\\
    \caption{Transverse momentum of isolated photon across different masses (left) and efficiency due to requiring photon $p_T > 10$ GeV (right) }
    \label{fig:photon}
    
\end{figure}

In the case of the isolated photon, its transverse momentum ($p^\gamma_T$) is closely linked to the mass of the light state ($m_\Phi$). 
Since the mass is a free parameter, all values in the range $2m_b<m_\Phi<m_Z$ are, in principle, allowed. The left plot of Fig.~\ref{fig:photon} demonstrates that an increase in the mass of $\Phi$ results in a systematic reduction of the transverse momentum of the associated photon. Thus, with  $m_\Phi\sim m_Z$, identification of the signal photon is a challenge owing to contamination from $\pi^0\rightarrow\gamma\gamma$ and from softer initial state radiation  radiation (ISR)/ final state radiation (FSR) photons.  As a result, we choose a lower threshold of $p_T>10$ GeV for the candidate photon and restrict the primary analysis to the five benchmark masses $m_\Phi=15,25,45,60,75$ GeV. The right plot of Fig.~\ref{fig:photon} illustrates the threshold effect, manifested as a reduced efficiency for signal events featuring at least one isolated photon. In particular, a sharp drop is observed after $m_\Phi=60$ GeV and is due to the increased number of candidate photons that fail this threshold. In general, with increasing mass of $\Phi$, the distribution for $p^\gamma_T$ becomes increasingly background-like.
 
\begin{figure}[htb!]
\includegraphics[width=0.5\linewidth]{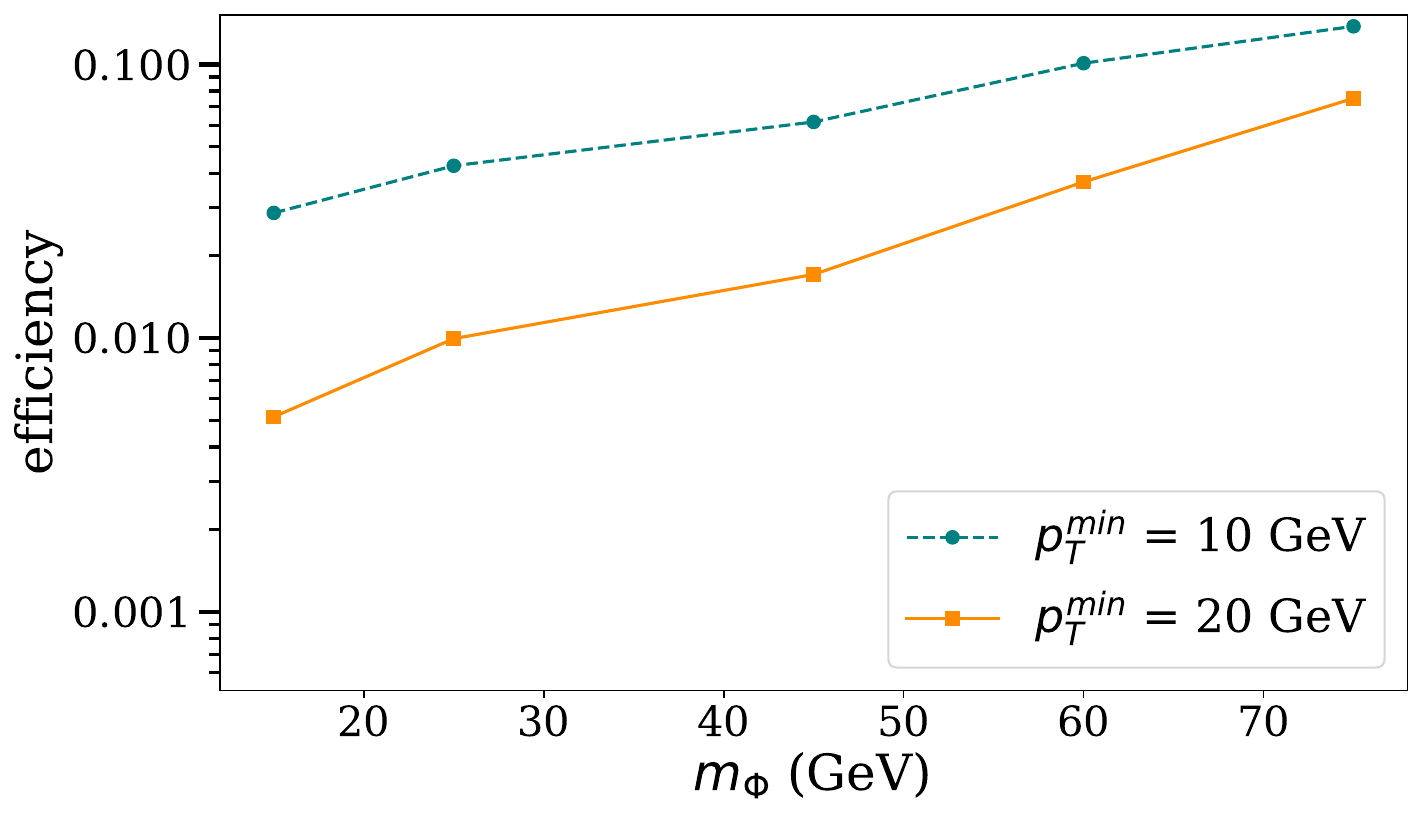} 
\includegraphics[width=0.5\linewidth]{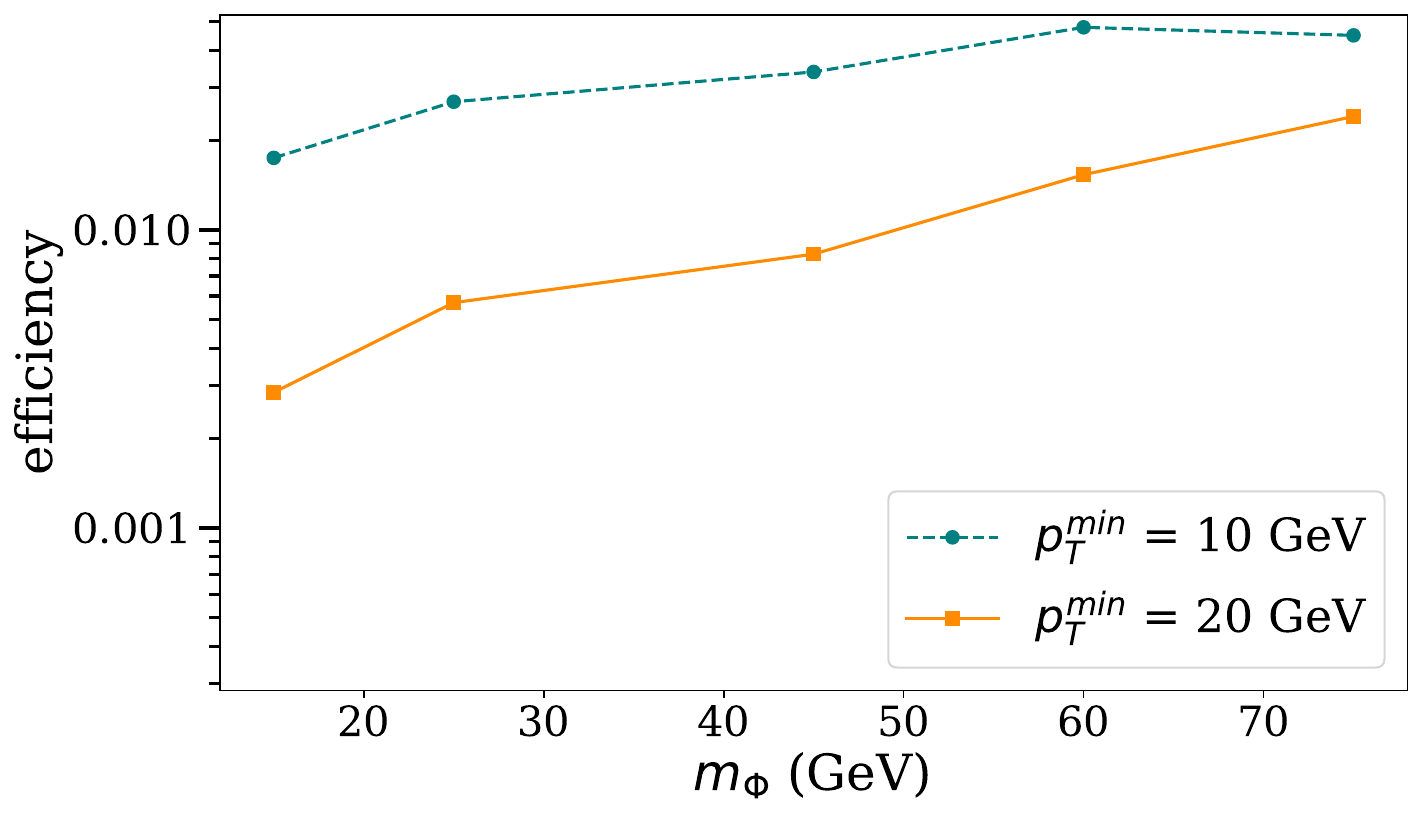}
  \caption{Efficiencies of the events with at least 2 b-tagged jets (left), at least 2 b-tagged jets and at least 1 photon (right) }
    \label{fig:bjetsefficiencies}
\end{figure}

Moving to jets, an important parameter in the reconstruction of the jet is the choice of the minimum transverse momentum, $p^{min}_T$ of the jet. This choice is driven only by the mass of $\Phi$ and hence the transverse momentum of its decay products. 
Left plot of Fig. \ref{fig:bjetsefficiencies} gives the signal efficiency, as a function of $m_\Phi$, for at least two $b$-tagged jets.  It is computed for two different values of $p^{min}_T$ for jet clustering: 20 GeV(orange) and 10 GeV(green). For a given $m_\Phi,$ a hierarchy in efficiencies between $p^{min}_T=10$ and $20$ GeV is observed. The effect is more pronounced for lower values of $m_\Phi$. This is due to the fact that for a given mass, the distributions of the transverse momentum of any of the outgoing $b$-quarks roughly peak at $m_\Phi/2$ while exhibiting a statistical spread about this value. Thus, the $ b$-quarks distributed in the lower end of this $p_T$ spectrum
are likely to be missed with a higher choice of $p^{min}_T$. In particular, choosing $p^{min}_T=20$ GeV for $m_\Phi=15$ GeV results in a loss of signal events as reflected in the efficiencies. 
With an increase in the mass of $\Phi$, the $p_T$ spectrum of the $b$-quarks shifts to higher values and consequently leads to an asymptotic convergence between $p^{min}_T=10$ and $20$ GeV.

 The individual analysis with the jets and the photon lends a better understanding of the observations for the overall event selection criteria: two $b$-jets and an isolated photon.
An interesting pattern is observed after $m_\Phi=60$ GeV, where a tendency for saturation in the efficiency is exhibited. This can be reconciled by the knowledge that a sharp drop in isolated photon efficiency at this mass is compensated by a corresponding increase in efficiency of at least two-$b$, thus exhibiting a saturation mechanism.  
Irrespective of this decrease, this analysis clearly indicates the $p_T^{min}=10$ GeV being the ideal choice that can optimise the overall signal acceptance across all the benchmark masses.
As already noted, there are two major sources of irreducible background: the di-jet QCD and the $Z\rightarrow b\bar b$ with the $\gamma$ radiated from one of the $ b$-quarks.
The corresponding efficiency for atleast two  $bb\gamma$ and $jj\gamma$  is observed to be $0.023(0.009)$ and $1.2\times 10^{-3}$ ($1.1\times 10^{-3}$), respectively, for $p^{min}_T=10(20)$ GeV.

The optimization of the event selection criteria is then followed by the identification of appropriate observables that could discriminate between the signal and the background. In addition to $p^\gamma_T$, another key feature of the signal topology is that the two $b$-quarks originate from a relatively `boosted' $\Phi$. This is particularly true when the mass $m_\Phi$ is significantly less than the mass of the $Z$-boson. In the limit where the $ b$ quarks are massless, the opening angle between them is given as $\Delta R\sim 2m_{\Phi}/p^\Phi_T$. 
\begin{figure}[htb!]
    \centering
     
    \includegraphics[width=0.48\linewidth]{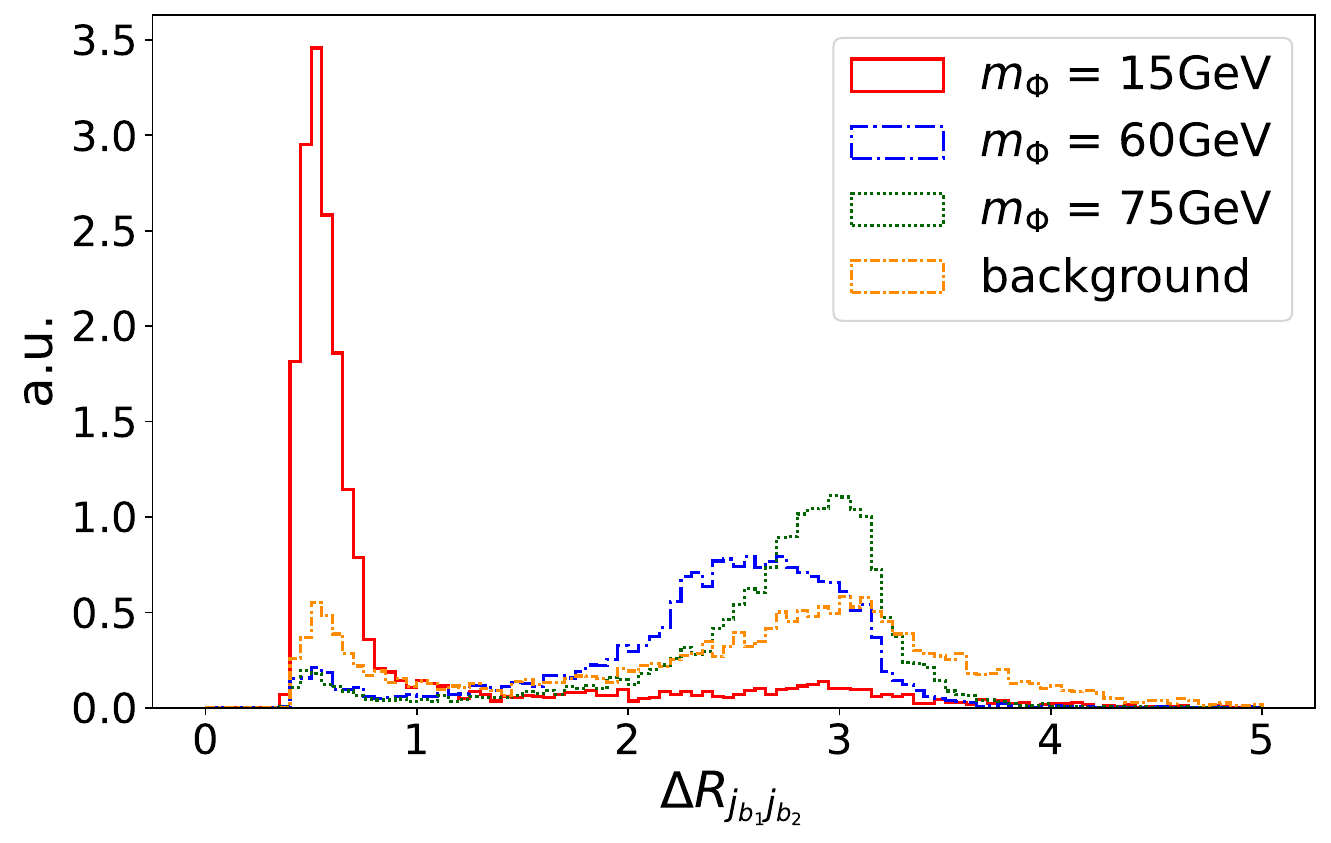}
       \includegraphics[width=0.48\linewidth]{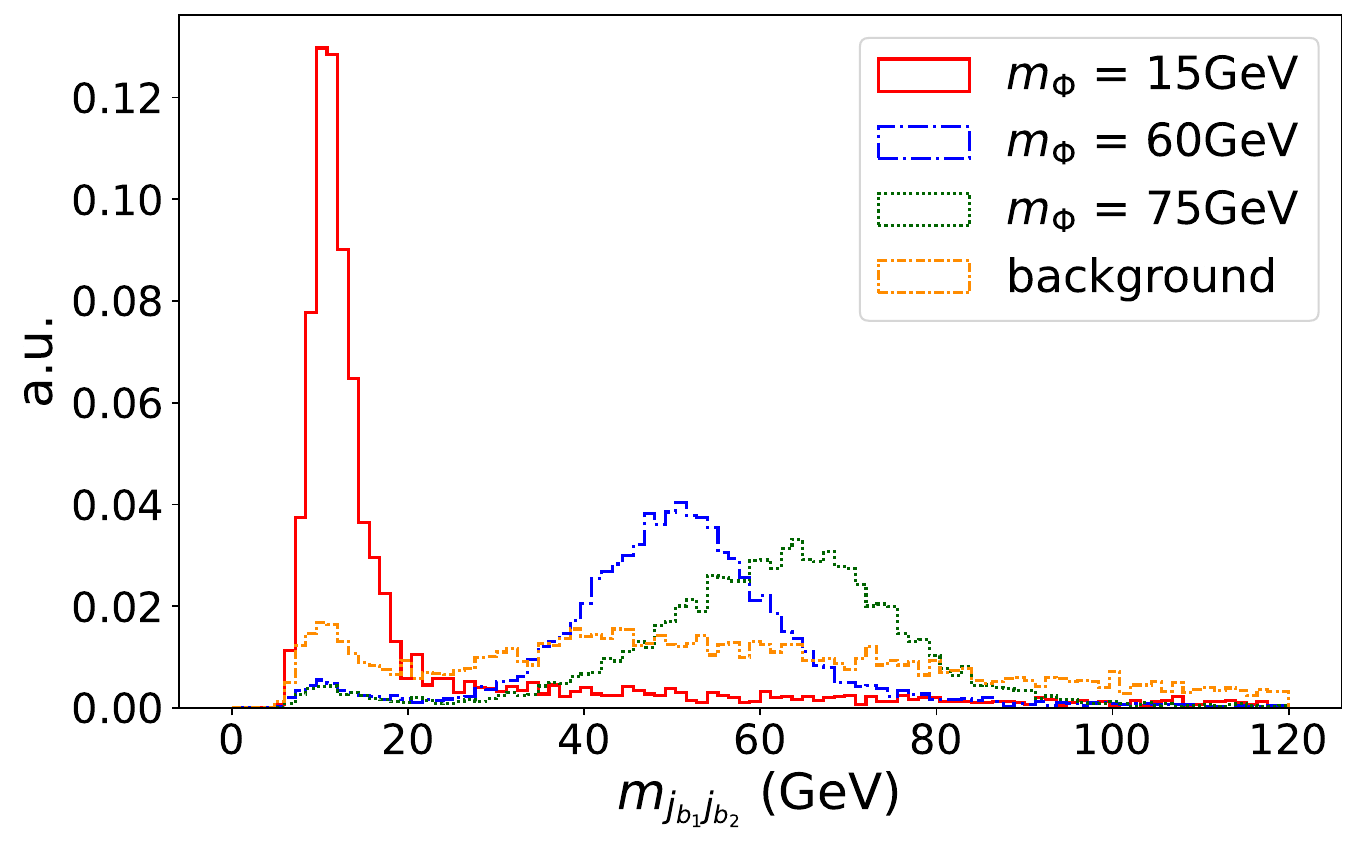}\\
       \includegraphics[width=0.48\linewidth]
    {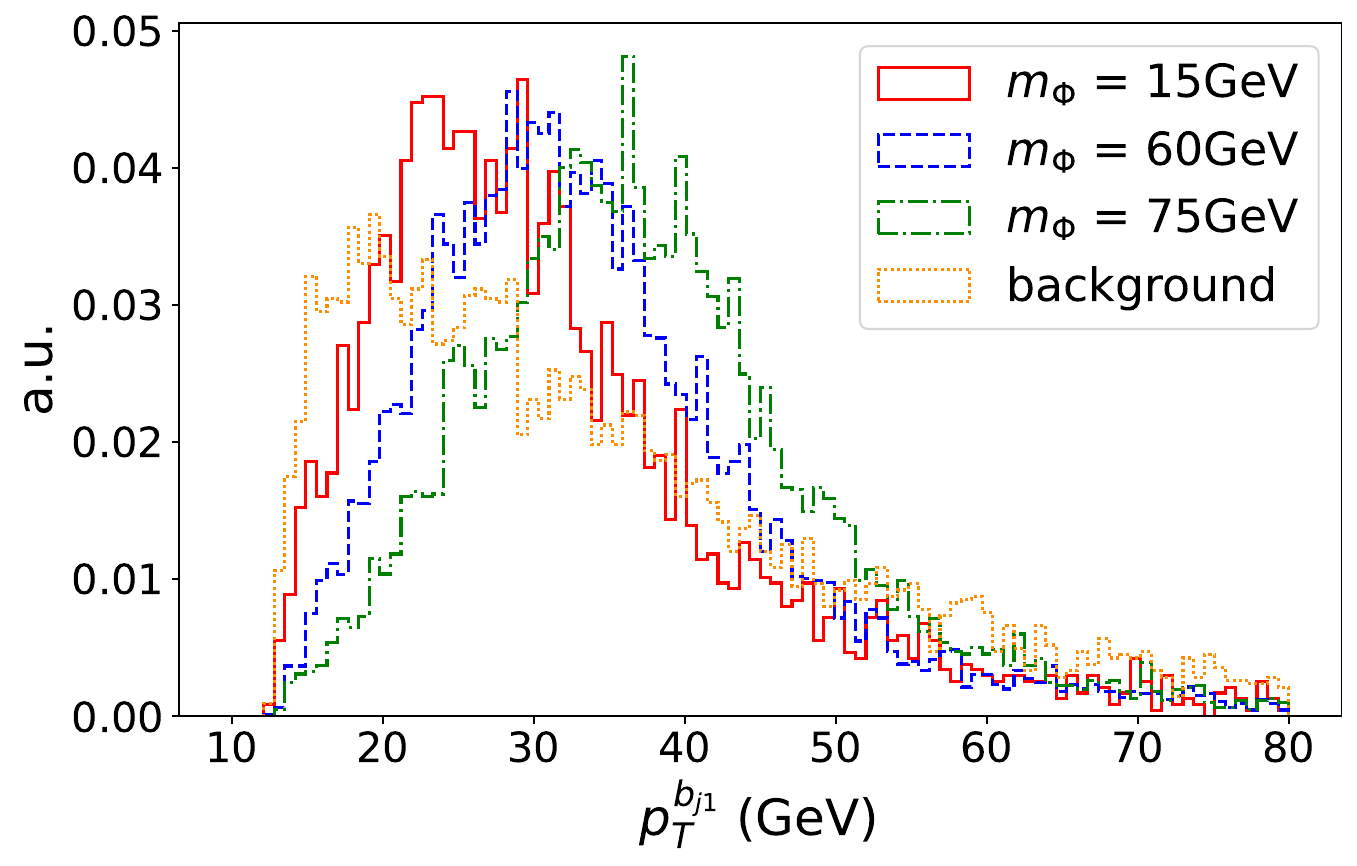}
    \includegraphics[width=0.48\linewidth]
    {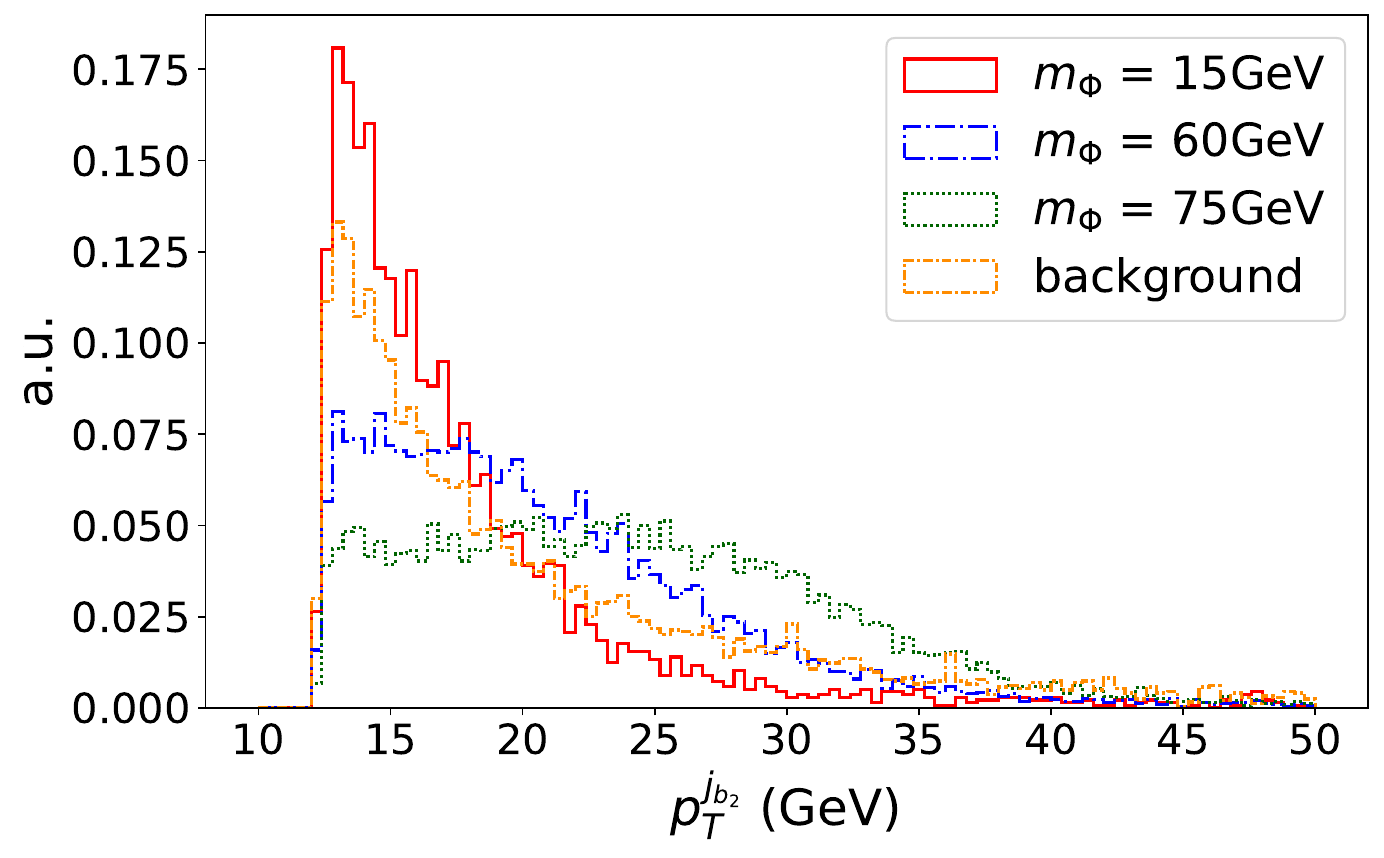}
    \caption{Distribution of the different variables for the background (orange) and different signal benchmarks }
    \label{fig:variables}
\end{figure}
The top left plot of Fig.~\ref{fig:variables} gives the computed $\Delta R_{j_{b_1}j_{b_2}}=\sqrt{\Delta \eta ^2+\Delta\phi^2}$ between the two $b$-tagged jets for different values of $m_\Phi$ and the background (orange). As expected, the background exhibits a higher value of $\Delta R_{j_{b_1}j_{b_2}}$  as they are produced with a large opening angle either from a $s$- channel process with an intermediate $Z,\gamma^*,g^*$ or a $t$-channel process. On the other hand, for the signal, $\Delta R_{j_{b_1}j_{b_2}}$ is roughly proportional to the value of $m_\Phi$.
For values of $m_{\Phi}$ closer to $m_Z$, the associated photon from the signal will be softer, and the $b$-quarks are likely to be further away due to the lower boost of $\Phi$.  Thus, both  $p_T^\gamma$  and $\Delta R_{j_{b_1}j_{b_2}}$   for the signal exhibit a similar pattern of being increasingly background-like with an increase in the mass of $\Phi$.

A defining feature of the signal is the presence of a light intermediate state.
As a result, its identification by means of its mass reconstruction is critical. The top right plot of Fig.~\ref{fig:variables} illustrates this reconstruction using the two $b$-tagged jets, ($m_{j_{b_1}j_{b_2}}$). 
The reconstruction for the background (orange) does not have a specific pattern as the extent of energy carried by the radiated photon of one of the $b$-quarks differs from one event to the other. A well-defined peak is visible for the signal, and the reconstruction becomes broader with increasing mass of $\Phi$. A possible explanation is that, for heavier masses and with $p^{\text{min}}_T = 10$ GeV used in jet reconstruction, the candidate $b$-jets may not originate from the signal. The bottom plots of Fig.~\ref{fig:variables} give the transverse momentum of the leading and the sub-leading $b$-jets. Its impact as a classifier is less pronounced for lower masses of $\Phi$, as much of the energy is carried away by the photon.

\section{Classification Schemes}
\label{sec:models}

\begin{figure}
    \centering
    \includegraphics[width=0.5\linewidth]{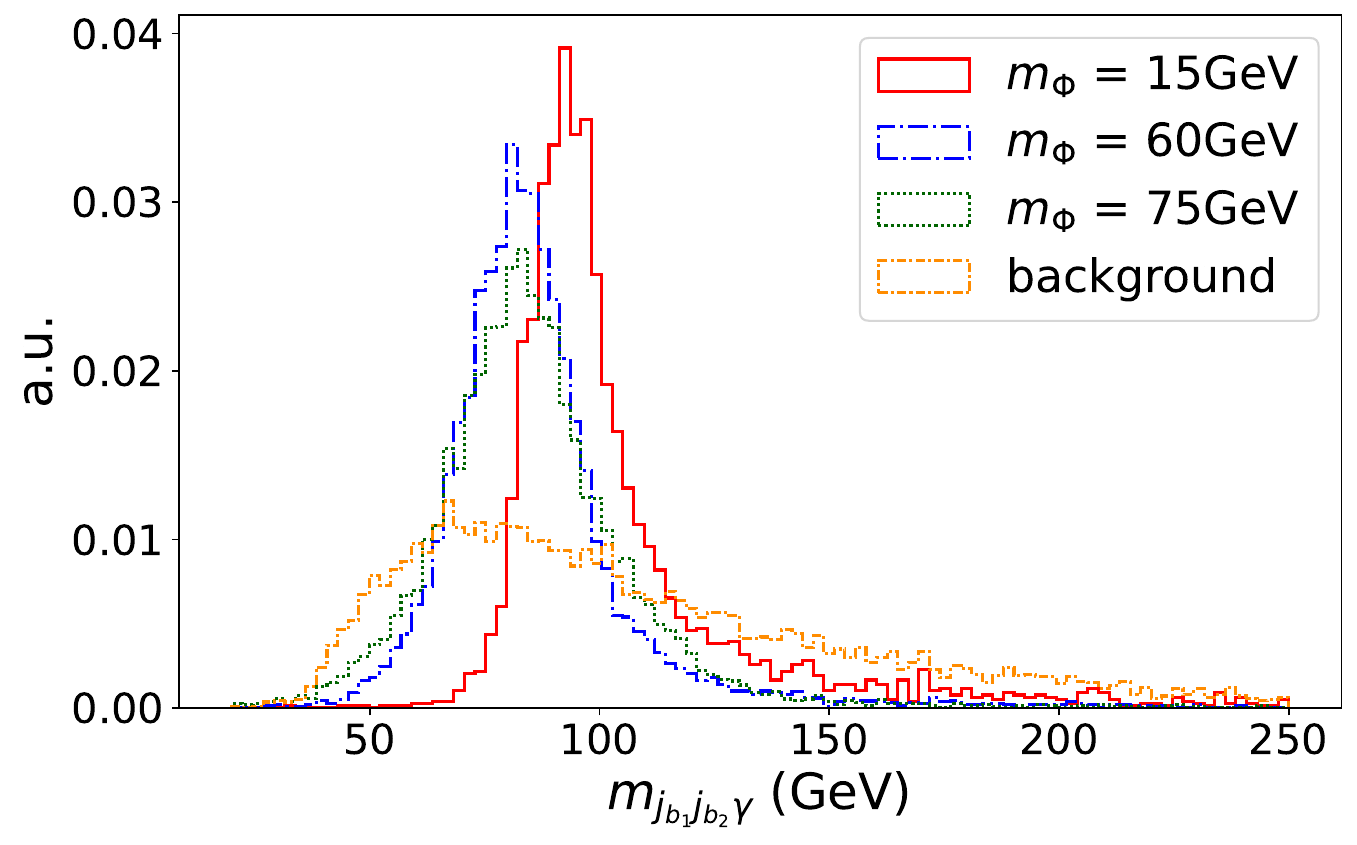}
    \caption{Invariant mass of two $b$-jets and an isolated photon}
    \label{fig:bbgamma}
\end{figure}

Identification of essential signal and background features paves the way for the implementation of an appropriate binary classification model. As the analysis attempts to explore the limits to which the HL-LHC is sensitive to this channel, we perform a comparative study of the sensitivity of different machine learning models to the intended signature. In Section ~\ref{sec:variables}, $\Delta R_{j_{b_1}j_{b_2}}$, $p_T^\gamma$, and $m_{j_{b_1}j_{b_2}}$ were identified as the key discriminating features across most values of $m_\Phi$, whereas $p_T^{j_{b_1}}$ and $p_T^{j_{b_2}}$ showed comparatively weaker discriminating power.
In addition, as we seek to identify the final state from a $Z$-portal, it is important to also consider the invariant mass of all the final states of interest:
$m_{j_{b_1}j_{b_2}\gamma}$ and is illustrated in Fig.~\ref{fig:bbgamma}.
The background has a relatively diffused reconstruction owing to the fact that the associated photon may not necessarily be from the final state radiation (FSR), as it may fail the isolation criteria. The photon candidate for the signal is more likely to pass the stringent isolation criteria.
Together with $m_{j_{b_1}j_{b_2}}$, the total invariant mass serves as a crucial indicator for identifying rare decays mediated through the $Z$-portal.

In general, all the variables fall under the category of low-level features, as only the four momenta of the $b$-tagged jets and the isolated photon are used in evaluating them. This leads to the consideration of the following two broad possibilities for the classifiers\footnote{Convolution neural Networks are ineffective in comparison as the analyses only utilises low level features and consequently does not result in well defined images. }: \\
A) Boosted Decision Trees ({\tt BDT}): At each node, the data is split based on one of the input features, resulting in the formation of a tree-like pattern. These splits are applied recursively to eventually minimize the entropy. Boosting then combines many such trees to construct a strong classifier\cite{he2019gradientboostingmachinesurvey}.\\
B) Graph Neural Networks: The primary objects in the event are the $ b$-tagged jets and the isolated photon. Their placements in the $(\eta,\phi)$ plane constitute the formation of a graph $G \equiv (V, E)$, where the nodes $V$ are the jets that are labelled as either a $ b$-tagged $j_{b_1,b_2}$ or an isolated photon. Thus, the $i^{th}$ node is associated with a feature vector $h_i\equiv(\eta,\phi,{\tt LABEL})$, where the {\tt LABEL} is $1$, if the node represents a $b$-tagged jet and $0$ if it is an isolated photon. 
The edges $E$ are the connections between the nodes with an associated feature $\Delta R=\sqrt{\Delta \eta^2+\Delta\phi^2}$. The graph network utilizes the message passing technique under which each node learns about its adjacent nodes. The extent of learning is broadly dependent on two aspects: 1) the existence of an edge between the two nodes, in the absence of which there is no message passing between them. For this analysis, we consider a fully connected graph that implies a well-defined edge between any pair of nodes. 2) The second aspect deals with the features of the nodes and the corresponding edges. Each node feature can undergo several consecutive stages of updates, where each stage is called a `layer'. 

\begin{figure}[htb!]
    \centering
    \includegraphics[width=15cm, height=3cm]{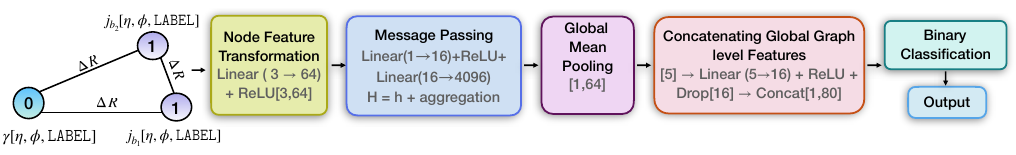}

    \caption{Architecture of the Graph Neural Networks used}
    \label{fig:GNN_architecture}
\end{figure}
The different schemes for message passing that are implemented for the analysis are summarized below:
\begin{itemize}
    \item Graph Convolution Network (\textbf{\tt GCNConv}): The node features are updated using the features of the neighbouring nodes that are connected by an edge. The update is weighed by a learnable weight matrix $\mathbf{\mathcal{W}}$. The $(l+1)^{th}$ update of the $i^{th}$ node, using the  {\tt GCNConv} operation can be mathematically expressed as \cite{GCNConv}
     \begin{equation}
    \mathbf{h}_i^{(l+1)} = \sigma \left( \mathbf{\mathcal{W}}^{(l)} \sum_{j \in \mathcal{N}_i \cup \{i\}} \frac{1}{\sqrt{\deg(i) \deg(j)}} \mathbf{h}_j^{(l)} \right)\
    \label{eq:gcnconv}
\end{equation} 
The degree of the $i^{th}$ node, $\deg(i)$, is the number of edges connected to that node, and $\sigma$ is the Rectified Linear Unit ({\tt ReLU}) operation \cite{nair2010relu}.
\item Neural Network Convolution (\textbf{\tt NNConv}): In this class of convolutional networks, the $i^{th}$ node feature in the $(l+1)^{th}$ layer is updated using a combination of edge features, as well as the node features in the $l^{th}$ layer, by means of a learnable weight metric $\mathbf{\mathcal{W}}$. The mathematical operation for the message passing is given as\cite{NNConv}

\begin{equation}
    \mathbf{h}_i^{(l+1)} = \mathbf{\mathcal{W}}^{(l)} \mathbf{h}_i^{(l)} + \sum_{j \in \mathcal{N}(i)} \mathbf{h}_j^{(l)} \cdot \mathcal{M}_{\mathbf{\mathcal{W}}}(e_{ij})\
\end{equation}
Here $\mathcal{M}:\mathbb{R} \to \mathbb{R}^{3\times3}\ $ is a multi-layer perceptron (MLP) that maps the edge features $e_{ij}\equiv\Delta R_{ij}$ to weights. A key feature that distinguishes this model from {\tt GCNConv} is that, in addition to updating the node features using a learnable weight matrix and the features of the neighbouring nodes, it also implements an MLP at each edge to generate edge-specific weight matrices.
 
\item Lorentz Equivariant Network ({\tt{LorentzNet}}): This is similar to the (\textbf{\tt NNConv}) method, where the node features are updated using the following expression \cite{LorentzNet}:
\begin{equation}
    \mathbf{h}_i^{l+1} = \mathbf{h}_i^l + c \sum_{j} \phi_x(m_{ij}^l) \cdot \mathbf{h}_j^l,
\end{equation}
where $m^l_{ij}$ denotes the output of an MLP constructed out of Lorentz scalars, $\phi(.)$ is another MLP, and $c$ is an appropriately chosen hyperparameter. A useful feature of {\tt LorentzNet} is its implementation of a Lorentz-equivariant message passing scheme, ensuring that the network inherently respects the symmetries of the Lorentz group. This is facilitated by the use of a combination of four-velocity vectors and Lorentz scalars as node features and a message passing mechanism that preserves Lorentz invariance.

\item Graph Attention Networks (\textbf{\tt GATConv}): The method employs an attention-based aggregation mechanism, wherein the interaction between nodes and edges is weighted by an attention coefficient, defined as \cite{GAT}
\begin{equation}
    \alpha_{ij} = \frac{\exp\left(\text{LeakyReLU}\left(\mathbf{a}^T [\mathbf{W} \mathbf{h}_i \Vert \mathbf{W} \mathbf{h}_j]\right)\right)}{\sum_{k \in \mathcal{N}_i} \exp\left(\text{LeakyReLU}\left(\mathbf{a}^T [\mathbf{\mathcal{W}} \mathbf{h}_i \Vert \mathbf{\mathcal{W}} \mathbf{h}_k]\right)\right)}\
\end{equation}
where $\mathbf{\mathcal{W}}$ is the learnable weight matrix, $a$ is the attention vector and $||$ represents the concatenation operation.
\end{itemize}
Once the message passing is implemented, the feature vectors of the different nodes are concatenated with the global feature vector
\[
\mathbf{g} \equiv \big( m_{j_{b_1}j_{b_2}\gamma},\; m_{j_{b_1}j_{b_2}},\; p_T^\gamma,\; p_T^{j_{b_1}},\; p_T^{j_{b_2}} \big)
\] to configure a Deep Neural Network (DNN) and set up the required binary classification model. The broad schematic that explains the general architecture of the different GNNs implemented is illustrated in Fig.~\ref{fig:GNN_architecture}. 

All GNN models were implemented using PyTorch Geometric \cite{paszke2019pytorchimperativestylehighperformance} and trained with the Adam optimizer \cite{kingma2017adammethodstochasticoptimization}, employing a learning rate of
$10^{-3}$ and a weight decay of $10^{-4}$.
A batch size of $64$ was used. The dataset was partitioned into training and testing subsets in a $70:30$ ratio, stratified by class labels. To ensure reproducibility, a fixed random seed ($42$ for PyTorch, NumPy) was applied.

For all GNN architectures, the node features were projected into a $64$ dimensional latent space. The baseline {\tt{GCN model}} employed a single {\tt{GCNConv}} message-passing layer, while the {\tt{NNConv}} model used a single edge-conditioned convolutional layer with 64 hidden dimensions. The GAT model utilized two message passing layers: the first with four attention heads producing 256 intermediate features, followed by a single-head projection back to $64$ dimensions. The LorentzNet model incorporated four sequential message passing blocks, each with $64$ hidden dimensions.

In all models, graph-level embeddings were concatenated with global kinematic features, which were separately projected into a 16-dimensional space. Classification was performed using a multi-layer perceptron (MLP) 
 and {\tt{ReLU}} activations. To mitigate overfitting, dropout regularization was applied at two levels: 0.3 in the feature-projection layers and 0.5 in the classifier layers. Training was carried out for 40 epochs for the GCN, GAT and NNConv models, and 60 epochs for LorentzNet, using the binary cross-entropy loss function \cite{crossentropy}.

\begin{figure}[htb!]
     \centering
     \includegraphics[width=0.5\linewidth]{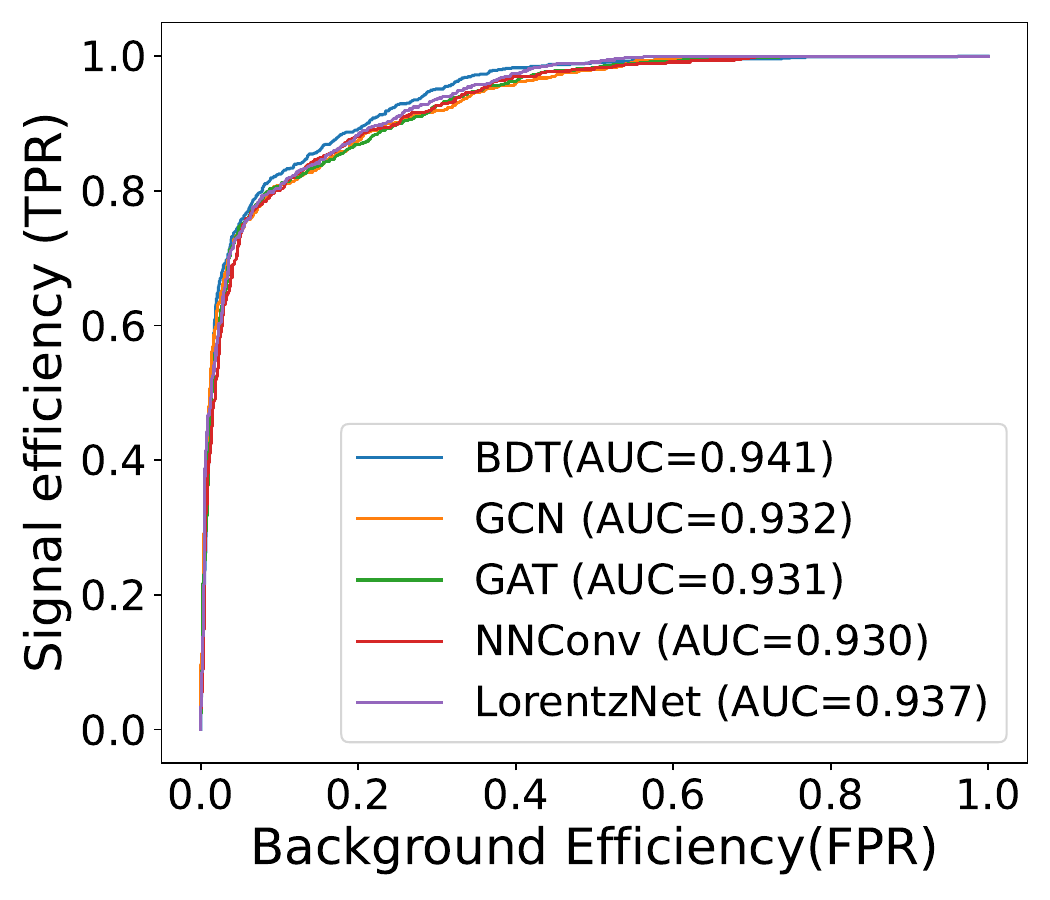}
     \caption{ROC curves for different architectures for $m_\Phi=15$ GeV.}
     \label{fig:ROCAUC}
 \end{figure}

For both the {\tt{BDT}} and {\tt{GNN}}, a classification threshold of
$\mathbf{T}= 0.5$ was used, where only events that satisfied $\mathbf{T}> 0.5$ were classified as signal. A discussion on the potential impact of changing a different threshold is presented in the Appendix \ref{app:threshold}.
A common metric for the comparison of the different architectures is the Receiver Operating Characteristic Area Under the Curve (ROC-AUC). Although a higher value for a particular model does not necessarily establish its superiority over the other frameworks, it serves as a useful guide for probing an architecture in more detail. Fig.~\ref{fig:ROCAUC} illustrates this comparison for $m_\Phi=15$ GeV by giving both the ROC curve and the corresponding AUC.
 Similar trends are noted for the other benchmark masses of $\Phi$ as well, thus demonstrating the stability of the results of classification across the diverse architectures. 
  
  In the case of {\tt BDT}, $\Delta R_{b_1b_2}$ emerged as the highest-ranked variable for all the signal benchmarks, indicating its dominant discriminating power. The same variable was processed slightly differently in a {\tt{GNN}}.
Working with low-level features and only three nodes, the different {\tt{GNN}} models implement a message passing scheme where the node features are updated using features of the connected nodes as well as the edge-features; $\Delta R_{b_1b_2}$. The node features are eventually concatenated with the global graph-level features, leading to the final classification. Without loss of generality, we adopt {\tt{BDT}} as the classifier of choice, and the ensuing discussion and results will be presented only for this scheme.

\begin{figure}[htb!]
    \centering
    \includegraphics[width=0.48\linewidth]{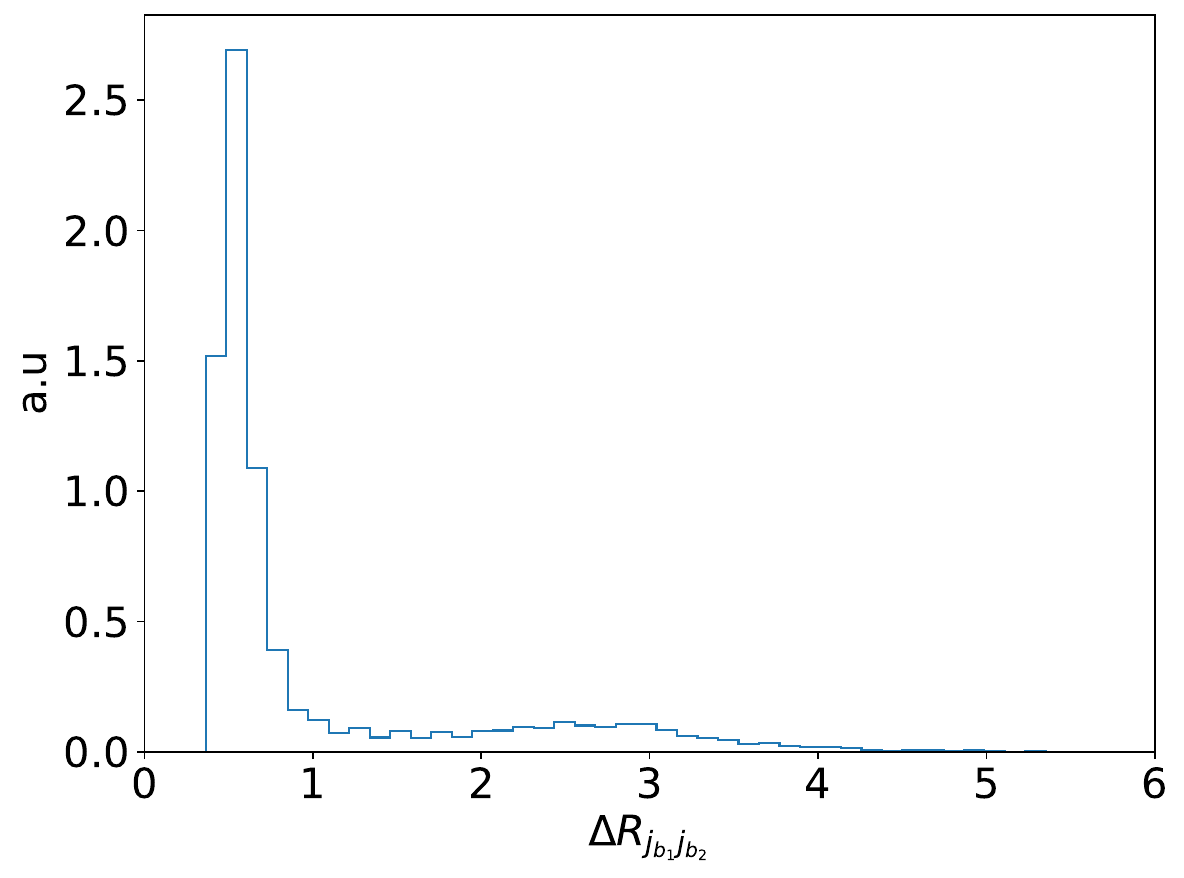}
    \includegraphics[width=0.48\linewidth]{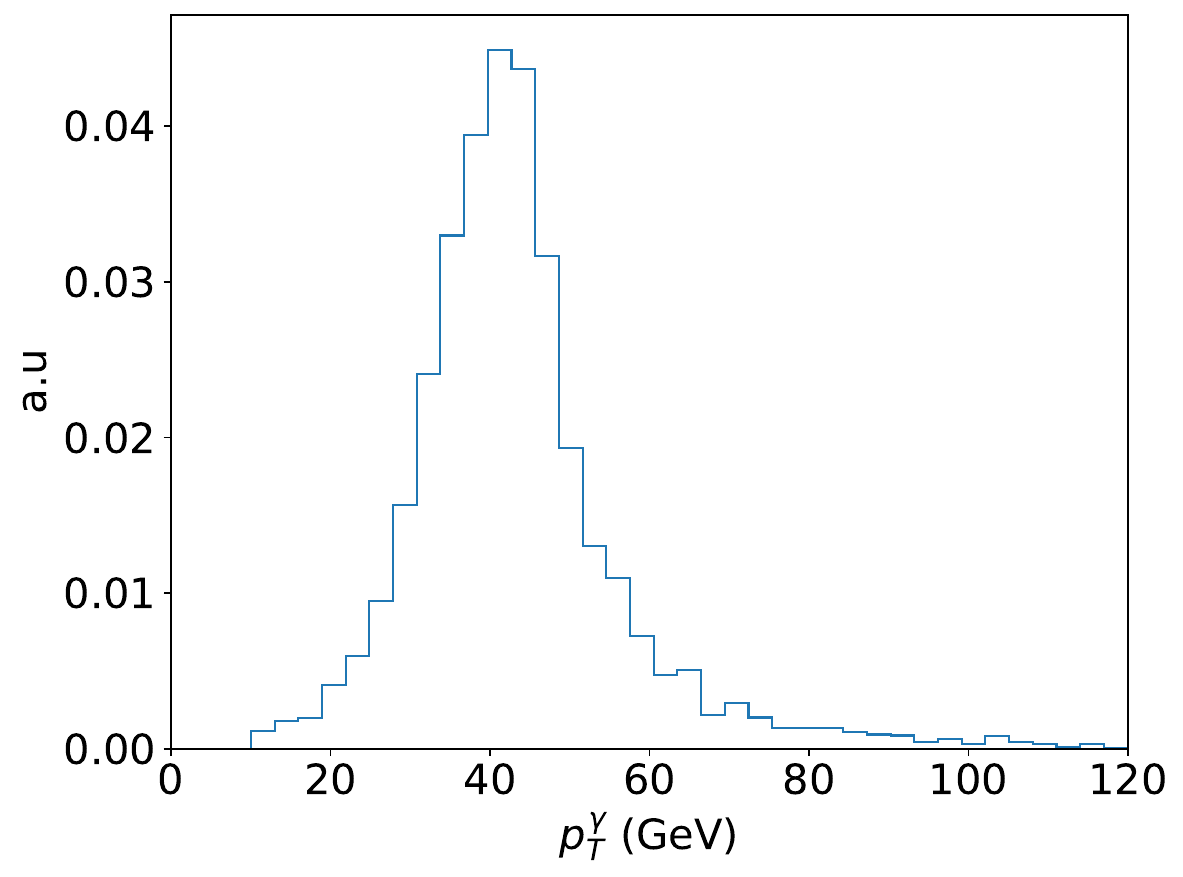} \\
    \includegraphics[width=0.48\linewidth]{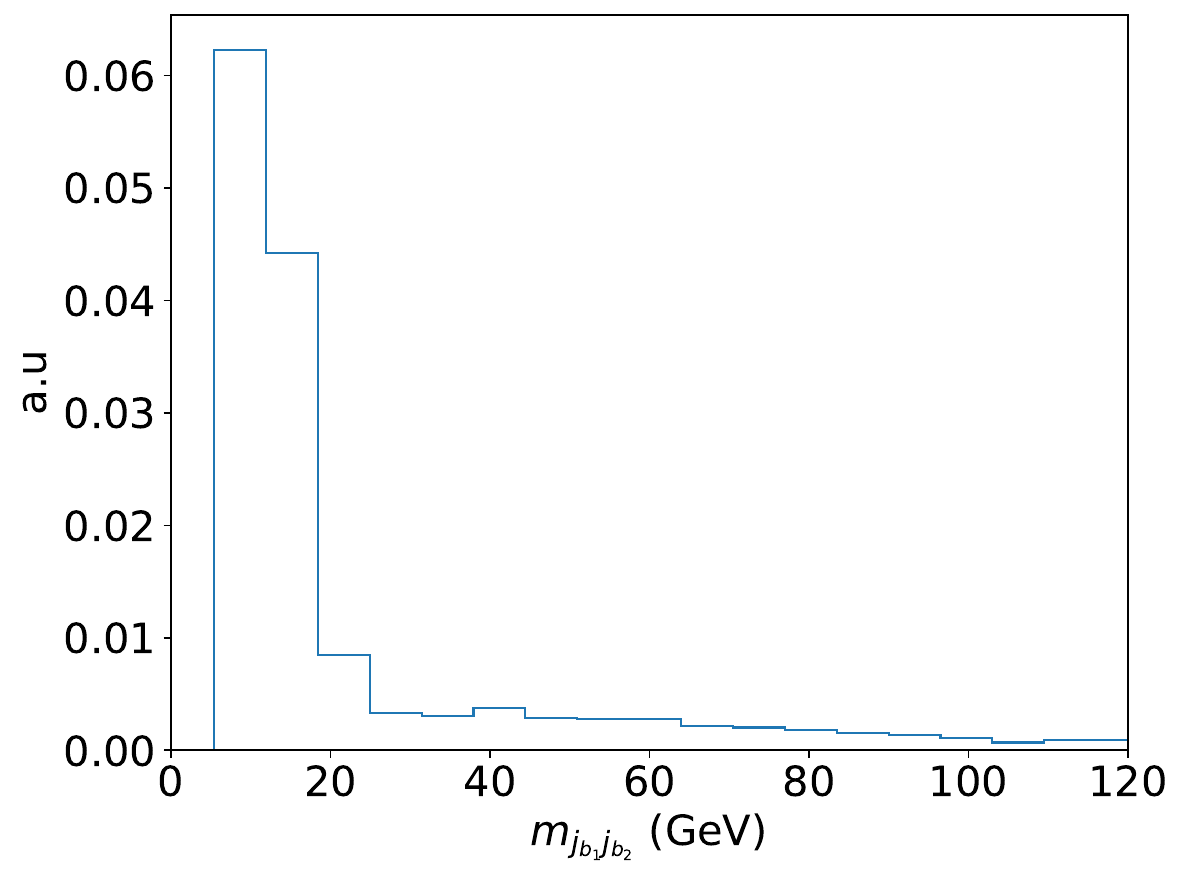}
    \includegraphics[width=0.48\linewidth]{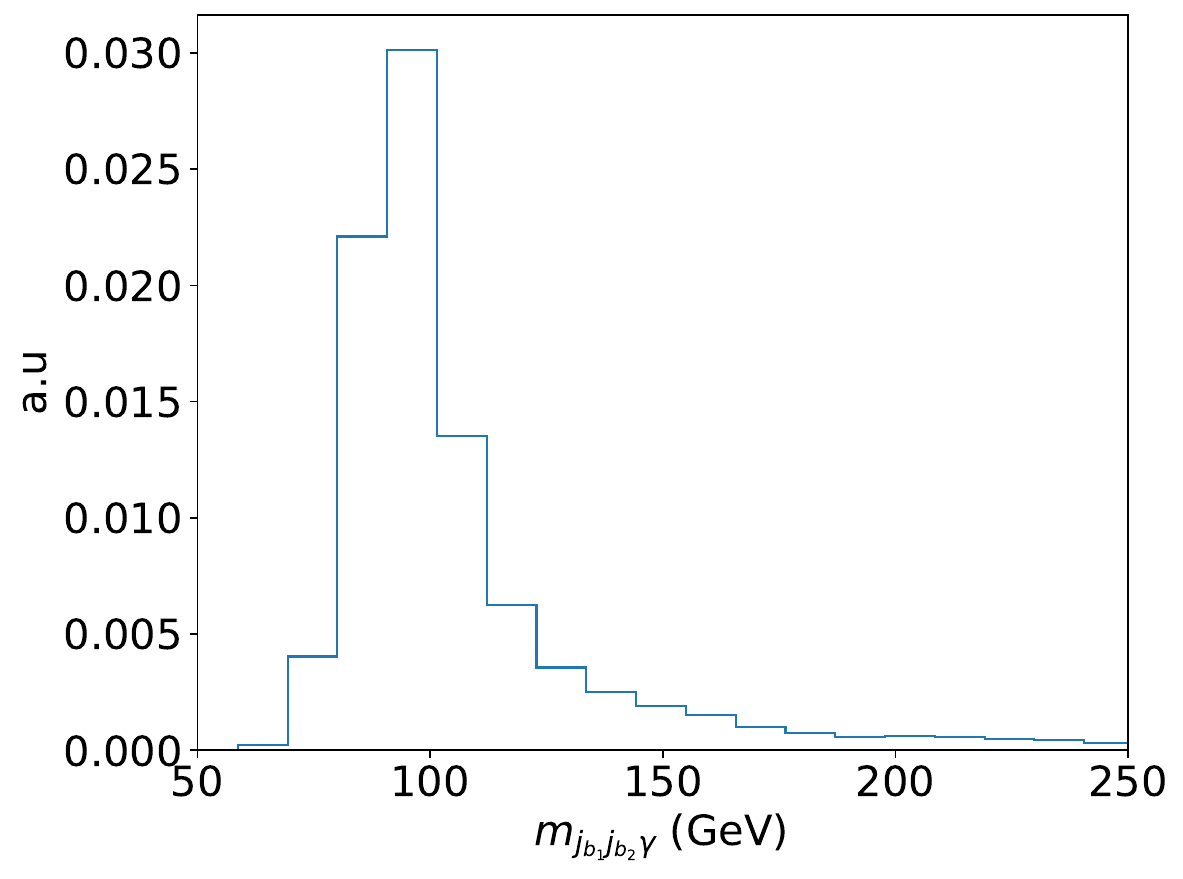}
    \caption{  Features from classification for the $m_\Phi=15$ GeV benchmark, for events crossing  threshold.}
    \label{fig:thresholdvariables}
\end{figure}

\begin{figure}[htb!]
    \centering
    \includegraphics[width=0.49\linewidth, height=6cm, keepaspectratio]{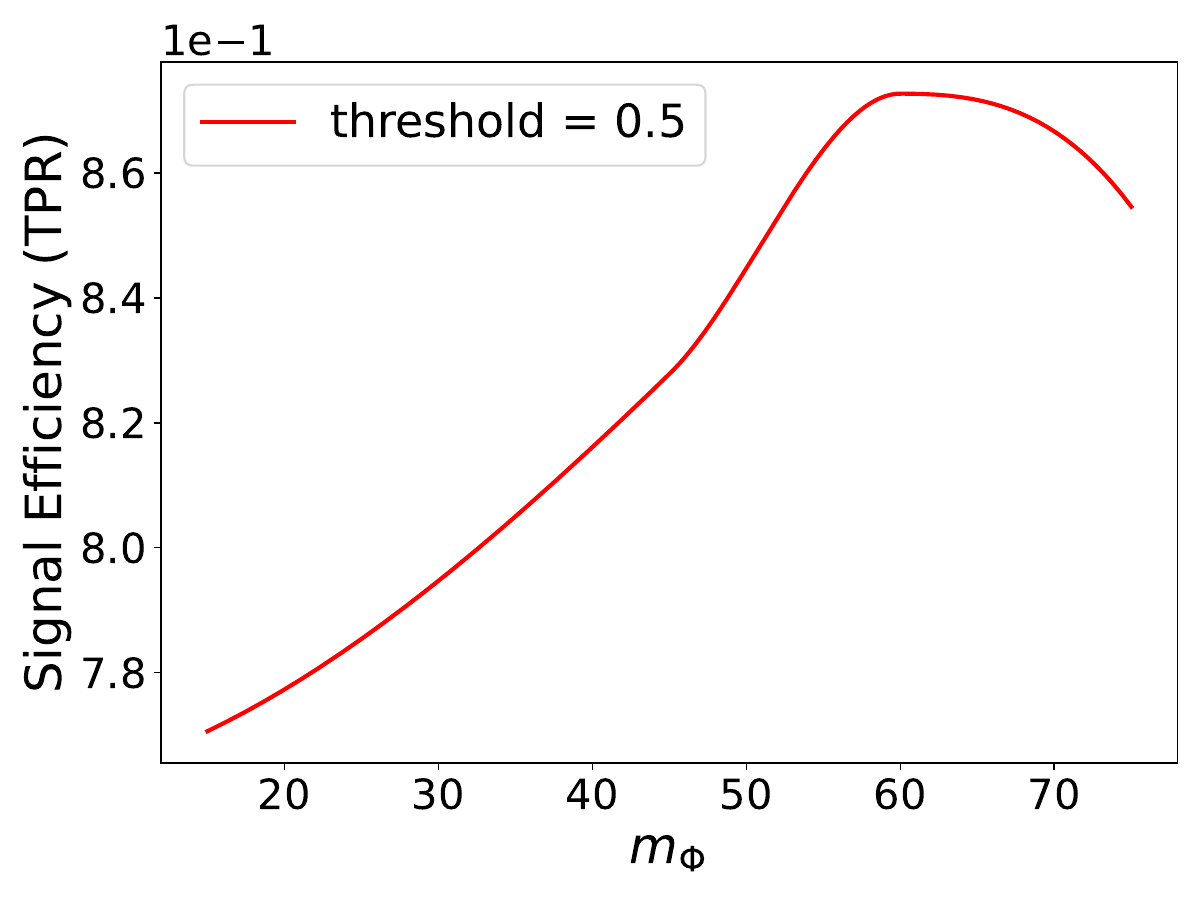}
    \includegraphics[width=0.49\linewidth, height=6cm, keepaspectratio]{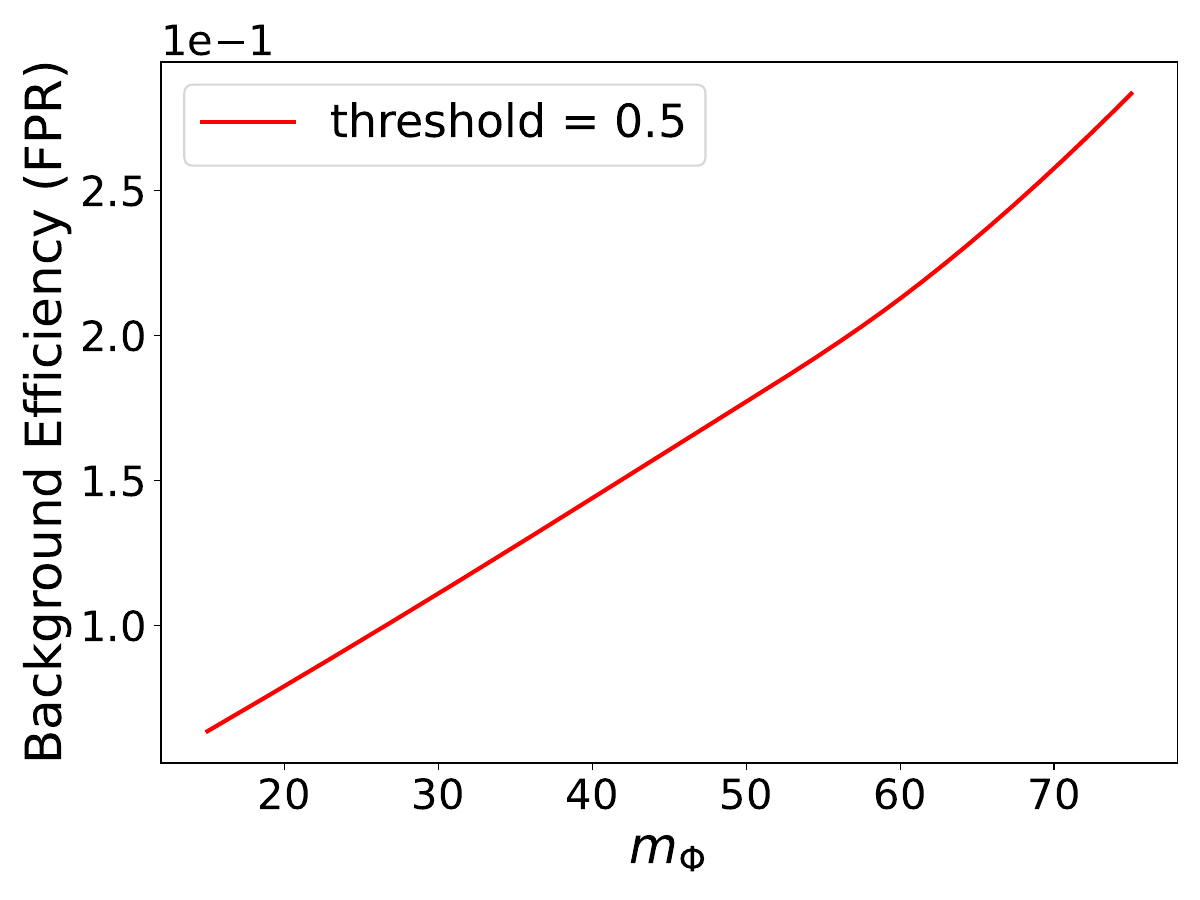}\caption{ True Positive Rate (left)and False Positive Rate (right) in \texttt{BDT} across different masses of $\Phi$.}
    \label{fig:tprfpr}
\end{figure}

\begin{figure}[htb!]
    \centering
    \includegraphics[width=0.48\linewidth]{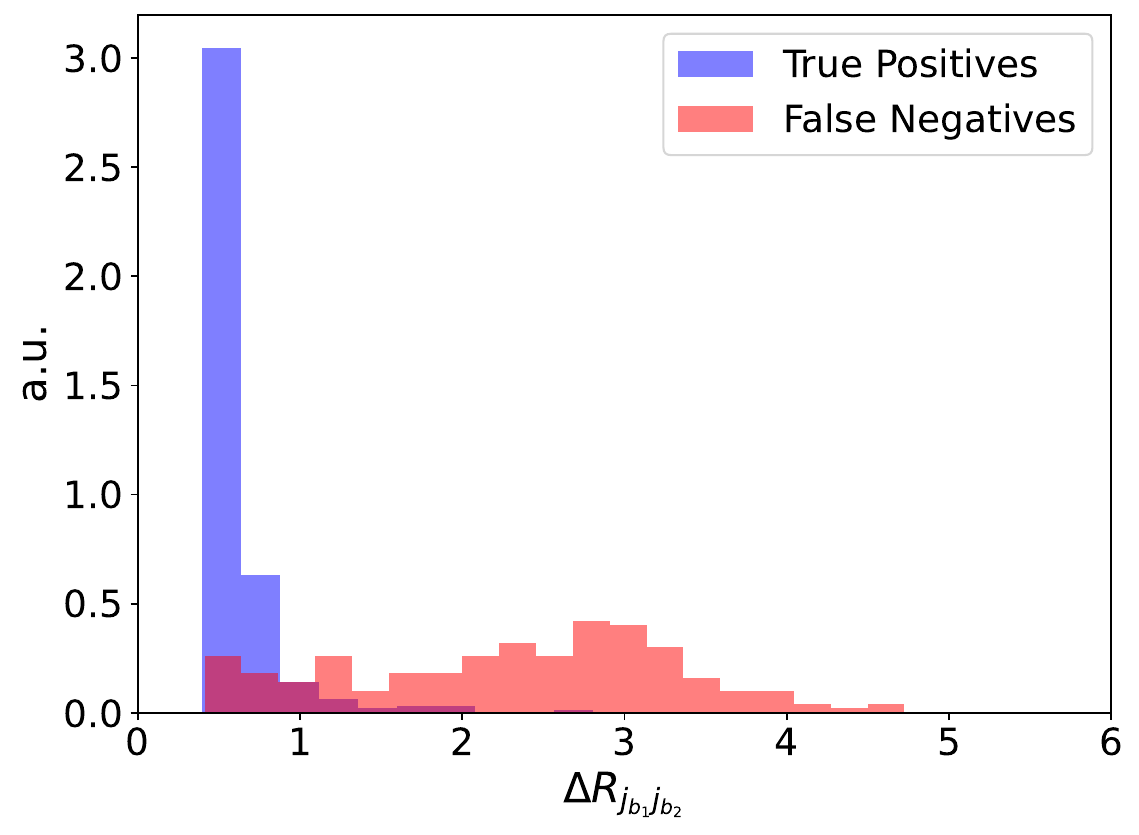}
    \includegraphics[width=0.48\linewidth]{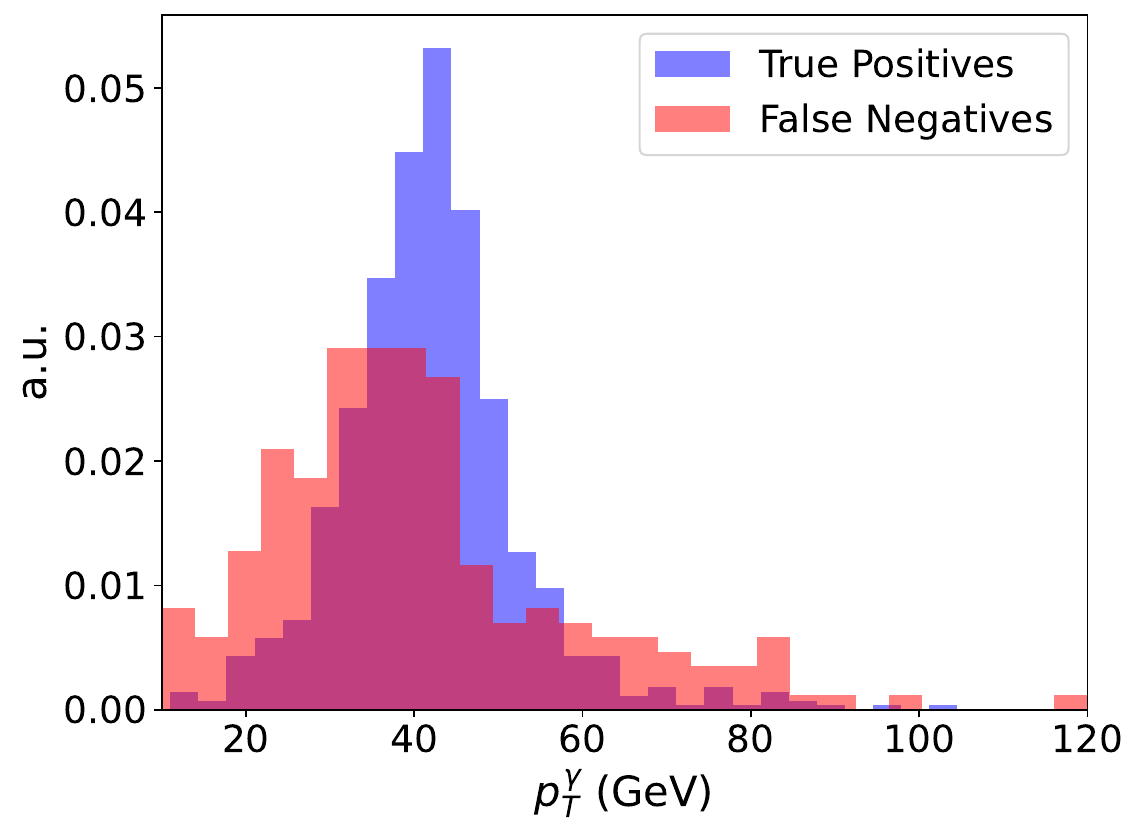} \\
    \includegraphics[width=0.48\linewidth]{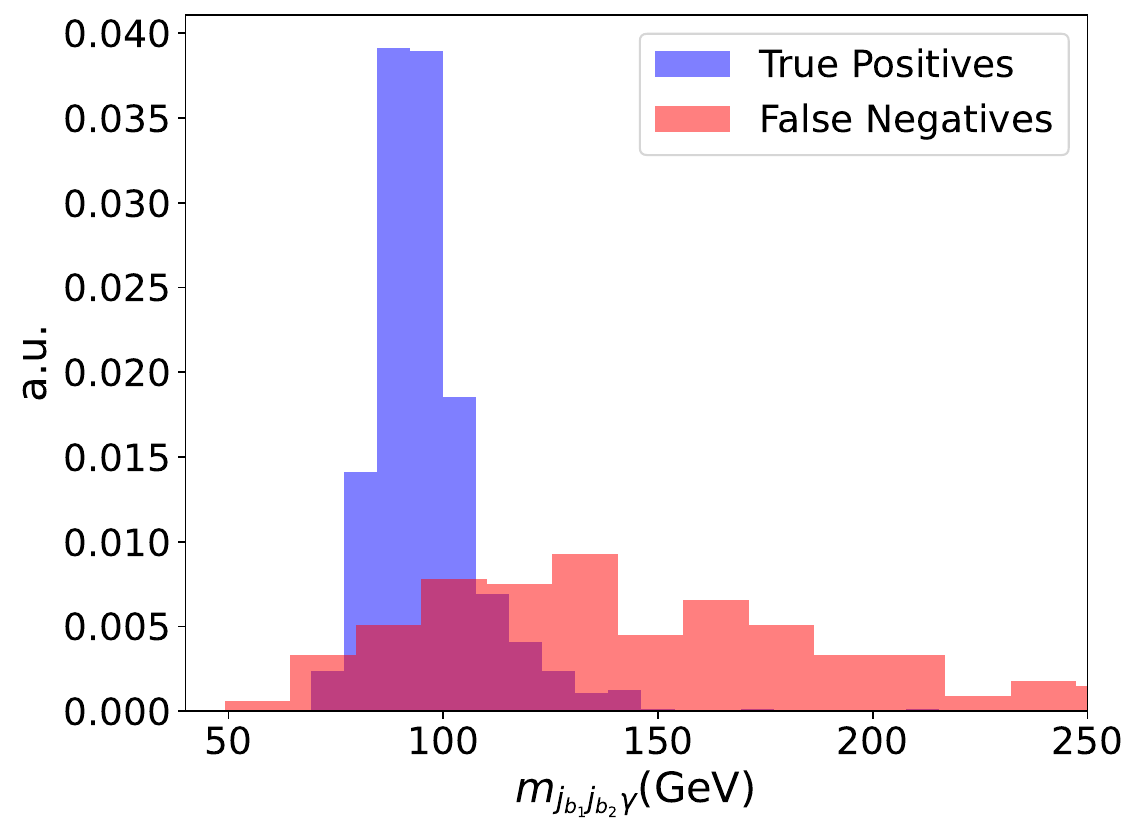}
    \includegraphics[width=0.48\linewidth]{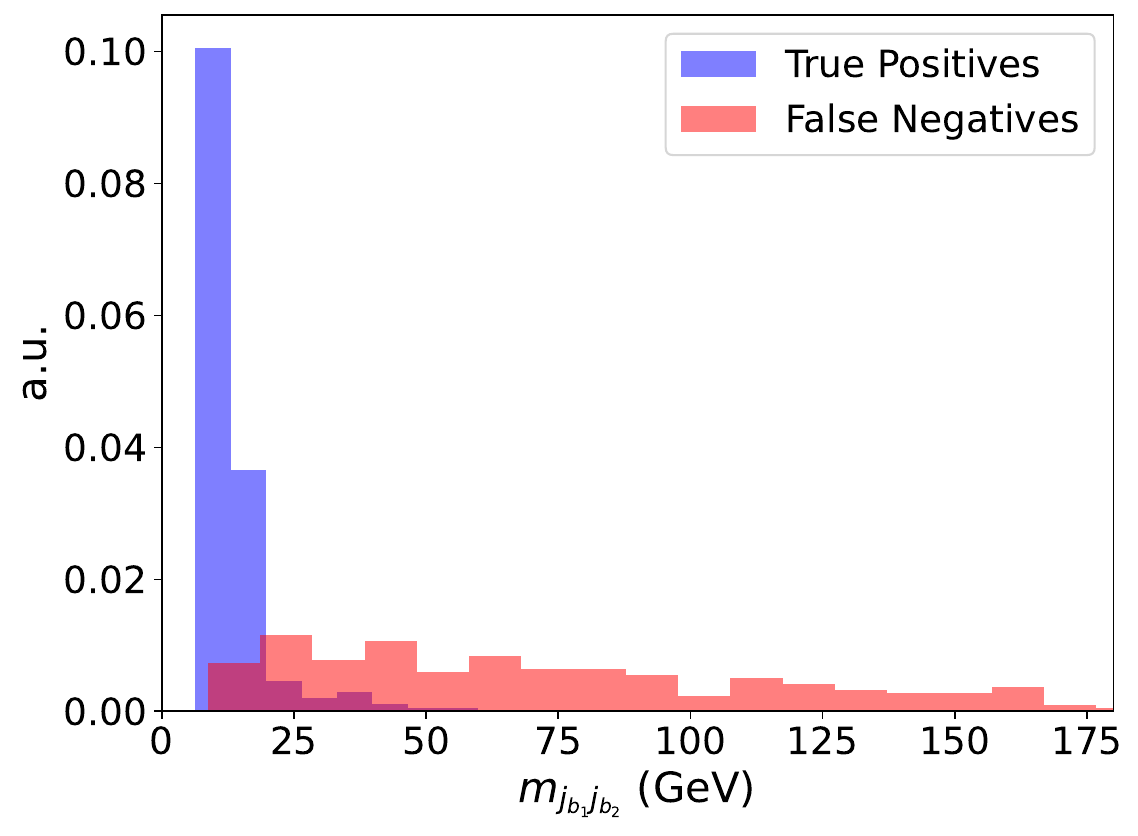} \\
    \includegraphics[width=0.48\linewidth]{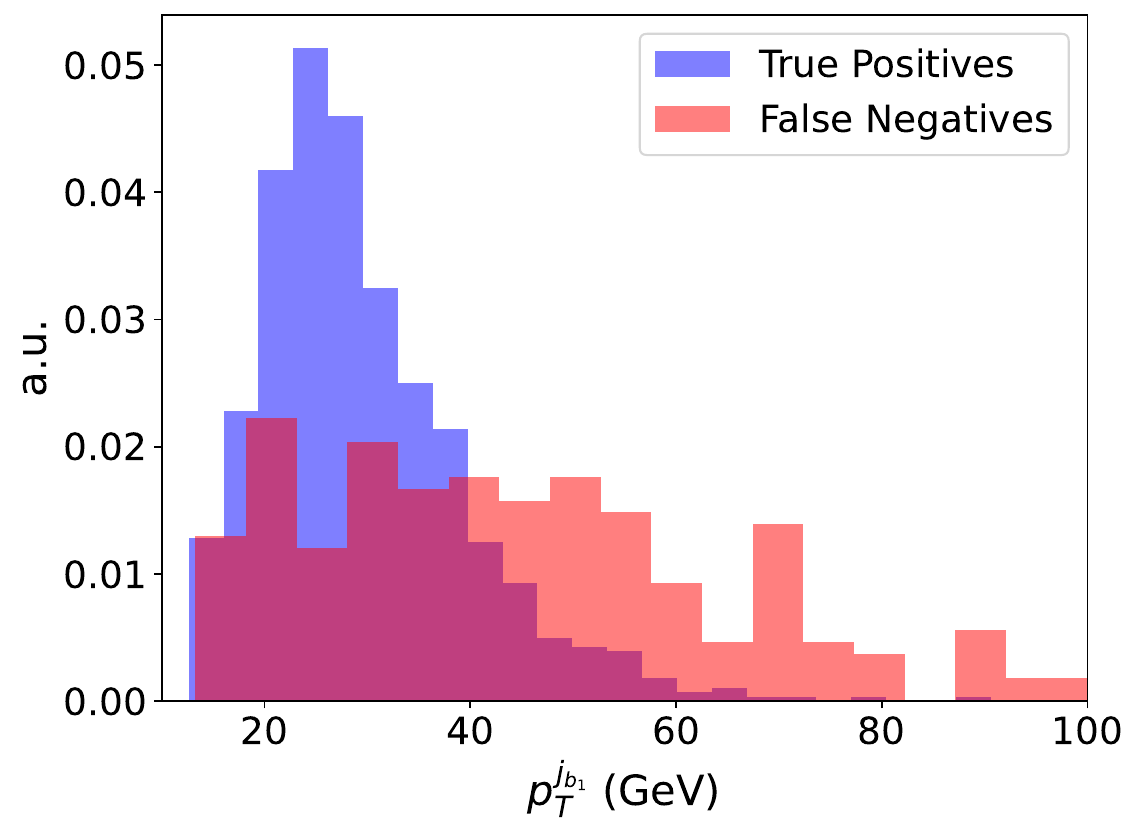}
    \includegraphics[width=0.48\linewidth]{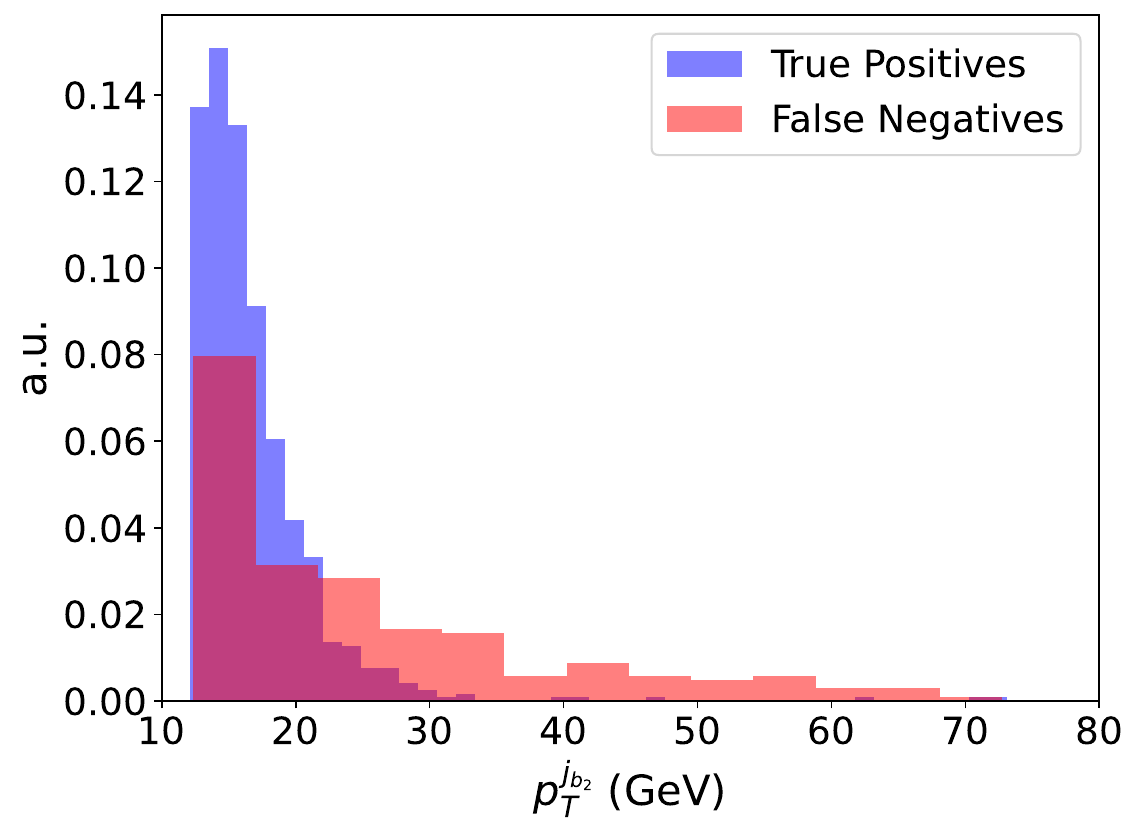}
    \caption{True Positives and False Negatives as predicted by the Gradient Boosting algorithm when $m_\Phi=15$GeV}
    \label{fig:GradientBoost_15GeV}
\end{figure}

Given the choice of a threshold,  it is interesting to note the distribution of the different kinematic features for events that are classified as signal by the {\tt{BDT}}. These are shown in \newline Fig. \ref{fig:thresholdvariables} for $m_\Phi=15$ GeV.  Visually, the individual distributions exhibit a striking similarity with the corresponding distribution in Fig. \ref{fig:variables}, where there appears to be little infiltration from the background events. 
A closer inspection, however, reveals interesting results.
The extent of signal events crossing and background infiltrating the threshold is quantified by the signal efficiency(True positive rate) and background efficiency(False positive rate), respectively. The left and right plots of Fig. \ref{fig:tprfpr} give the respective efficiency for different masses. The signal efficiency (left) is lowest for 15 GeV and monotonically increases until 60 GeV before falling again. The pattern may seem counterintuitive given that the key variables for the lower masses seemed to exhibit better discrimination from the background, as shown in Fig. \ref{fig:variables}. However, the extent of signal events that fail the threshold is also higher for lower mass as illustrated in Fig. \ref{fig:GradientBoost_15GeV} for $m_\Phi=15$ GeV. It compares the kinematic distributions for all the input features separately for events that cross the threshold (True positives) and those that fail (false negatives). It is to be noted that the effect is pronounced for all variables that involve either of the $b$-tagged jets. As the lower masses are associated with relatively low $p_T$ $b$-quarks, their probability of being identified as a $b$-jet is comparatively lower. As a result, the jet tagged as a $b$-jet may also be due to a gluon splitting instead from the decay of a $\Phi$.

The maximum value for the signal efficiency is observed for $m_\Phi=60$ GeV. Not only does this represent an optimal point that is well discriminated from the background, but also the signal events are more likely to cross the threshold, thus reducing the number of false negatives. While the $b$-jets from this benchmark are better identified, the smearing of their momenta reflects less than perfect reconstruction of the masses. This is one of the reasons why a signal point for this benchmark may be falsely labelled as a background. Fig. \ref{fig:GradientBoost_60GeV} in Appendix \ref{app:tprfnr60} illustrated the distribution of the True Positives and the false negatives for this benchmark. Beyond $m_\Phi=60$ GeV, there is a drop in the signal efficiency as the signal becomes increasingly background-like, thereby resulting in the reduction of true positives.

Another important variable in estimating the final signal sensitivity is the background efficiency or false positive rate (FPR).
The right plot of Fig.~\ref{fig:tprfpr} shows FPR as a function of $m_\Phi$. The monotonic increase is a consequence of the fact that the signal characteristics are becoming increasingly background-like, thus resulting in many background events crossing the threshold, thereby increasing the FPR.

\section{ Results and Discussions}
\label{sec:results}
The framework and the classification strategy discussed in Section \ref{sec:models} provide the necessary foundation to estimate the sensitivity of the HL-LHC to the $b\bar{b}\gamma$ decay mode of the $Z$-boson. A key aspect of the analysis was the adaptation of a universal strategy for all the benchmark masses for $\Phi$. While this does not rule out finer improvements by focusing on specific masses, the universal strategy illuminates the aspect that needs attention from the experimental perspective. 

 As we adopt a model-independent strategy, the goal would be to quantify the limits up to which the HL-LHC can probe the particular decay mode of $Z$. We consider the following log-likelihood ratio
\begin{equation}
    q=-2\log\left(\frac{L_{SM}}{L_{SM+S}}\right)
\end{equation}
where $SM(S)$ correspond to the Standard Model (Signal) hypothesis, respectively.
The sensitivity, $Z=\sqrt{q}$ is a measure of the extent of deviation from the SM hypothesis.
As already discussed, the precise measurements at the $Z$ pole will render this particular decay mode of the $Z$ to be extremely rare.
This makes the background significantly more dominant than the signal, thus simplifying the sensitivity $Z\sim S/\sqrt{B}$. Denoting B.R$(Z\rightarrow \gamma\Phi)=\alpha$, the signal (S) events at the end of the HL-LHC is
\begin{equation*}
    S=\sigma_Z\epsilon_{S}\alpha \mathcal{L}~\text{TPR}
\end{equation*}
where $\sigma_Z$ is the $Z$ boson production cross-section at 14 TeV,  $\epsilon_{S}$ is the acceptance efficiency for the signal after implementing the event selection criteria as shown in Fig. \ref{fig:bjetsefficiencies}, $\mathcal{L}=3~\text{ab}^{-1}$ is the integrated luminosity and TPR is the true positive rate after the implementation of the binary classification as shown in Fig. \ref{fig:tprfpr}.

Similarly, the background events $B$ are given by
\begin{equation*}
     B=\left(\sigma^{b\bar{b}\gamma}\epsilon_{b\bar{b}\gamma}+\sigma^{jj\gamma}\epsilon_{jj\gamma}\right) \mathcal{L}~\text{FPR}
\end{equation*}
where $\sigma^{k}\epsilon_{k}$ is the product of the background cross-section and event selection efficiency for $k=b\bar{b}\gamma,jj\gamma$ and FPR represents the false positive rate. The leading order cross-sections, obtained from {\tt{MADGRAPH}}, for the backgrounds are: $\sigma^{b\bar{b}\gamma}=1.9\times 10^{4}$ pb and $\sigma^{jj\gamma}=3.09\times 10^{5}$ pb. We assume a common value $1.5$ as \textit{k-}factor for both the backgrounds \cite{kim2024panoramicstudykfactors111}.

 \begin{figure}[htb!]
     \centering
     \includegraphics[width=0.6\linewidth]{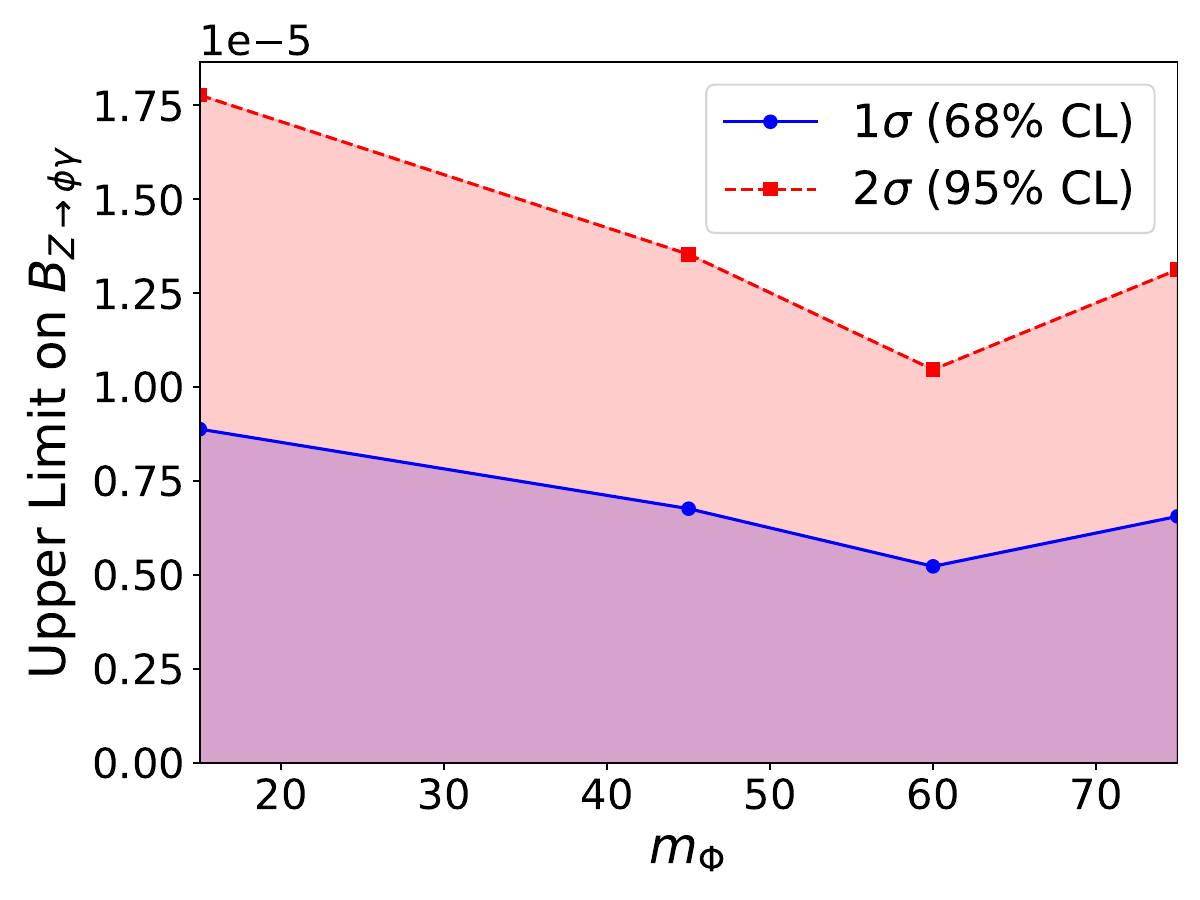}
     \caption{The estimated upperbound on the BR$(Z\rightarrow \gamma\Phi)$ as a function of $m_\Phi$ at 68$\%$ and 95$\%$ CL.}
     \label{fig:upperbound}
     \end{figure}

This can then be translated to obtain the upper bound on BR$(Z\rightarrow \gamma\Phi)$ as a function of the mass of $\Phi$ and illustrated in Fig.~\ref{fig:upperbound}. It is shown for the $68\%$ (blue) and $95\%$ (red) confidence intervals. The bound is strongest for $m_\Phi=60$ GeV, mirroring the pattern obtained for the TPR in Fig.~\ref{fig:tprfpr}. The upper bounds are particularly promising, where even the weakest bound corresponds to $10^{-5}$ for $m_\Phi\sim 15$ GeV.

The analysis has laid the foundation for identifying several necessary advancements, particularly for low $p_T$ $b$-jet tagging. This regime is beset by relatively low efficiencies and a comparatively larger number of false negatives. This can be largely overcome by improving the tagging efficiencies for low $p_T$ in the context of HL-LHC. While higher efficiencies are expected for future colliders like FCChh\cite{Selvaggi:2717698}, the study provides a strong motivation for pushing the limits of the HL-LHC.

\section{Conclusions}
The paper developed a concrete foundation for the search for rare $Z\rightarrow\Phi\gamma$ decay in the $b\bar b\gamma$ decay mode.  A universal event selection criteria, to be applied for all masses of $\Phi$, was identified and eventually passed to a universal binary classification scheme. 
A remarkable agreement between the results of {\tt{BDT}} and different {\tt{GNN}} architectures was noted, thus laying down two parallel paths for future improvements: A) The $b$-jets from the decay of the $\Phi$ were characterized by low $p_T$ and correspondingly low tagging efficiencies. This was particularly true for lower masses of $\Phi$, where the smaller signal efficiencies partially nullified the advantages enjoyed by the discriminating features for these benchmarks. An improvement in the low $p_T$ $b$- tagging efficiencies will have wide-ranging applications. In addition to its utility for the type of signal considered in the paper, it can also be extended for soft $b$-jet identification from cascade decays and typical in models with compressed spectrum.
B) Finer improvements are possible by employing separate event selection criteria as well as classification schemes that are tailored for individual masses. This will be particularly useful in the context of precision studies. It is expected that the proposed hadron colliders like FCC-hh\cite{Selvaggi:2717698} are expected to have better low-$p_T$ $b$-tagging. Thus, the eventual signal efficiencies can be improved by at least a factor of two, thereby leading to stronger constraints and potentially a discovery. As new physics continues to be increasingly elusive, the need of the hour is not only to develop advanced state-of-the-art classification architectures that are sensitive to even the smallest perturbations about the SM. It is also necessary to identify and give a closer look at kinematic regimes where unexpected phenomena can lie in store.

\section*{Acknowledgements}
We are thankful to Pankaj Borah, Abhishek Kumar Singh for a careful reading of the manuscript. AMI and TT acknowledge useful discussions with Sanmay Ganguly. AMI acknowledges the generous support by SERB India through project no. SRG/2022/001003. AMI and TT acknowledges Meet Jain and Kashish Jangra for their insightful discussions on the Graph Neural Networks. The research of TT is supported by the Institute Fellowship from the Indian Institute of Technology Delhi, India.


\providecommand{\href}[2]{#2}\begingroup\raggedright\endgroup

\newpage
\appendix

\section{Impact of choice of threshold}
\label{app:threshold}
\begin{figure}[H]
    \includegraphics[width=0.5\linewidth, height=0.25\textheight]{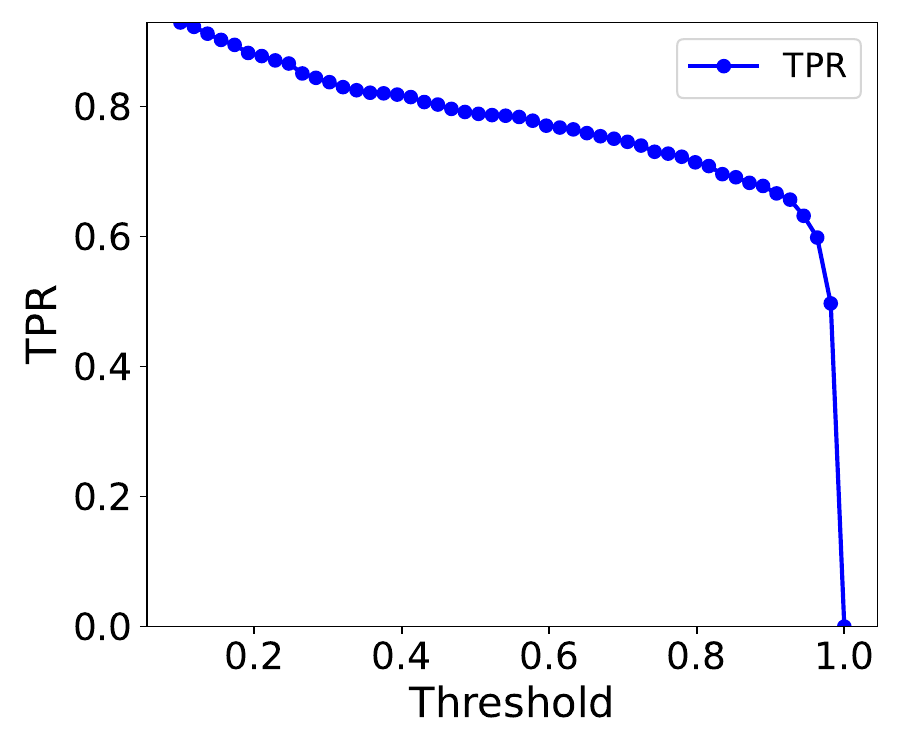}
    \includegraphics[width=0.5\linewidth, height=0.25\textheight]{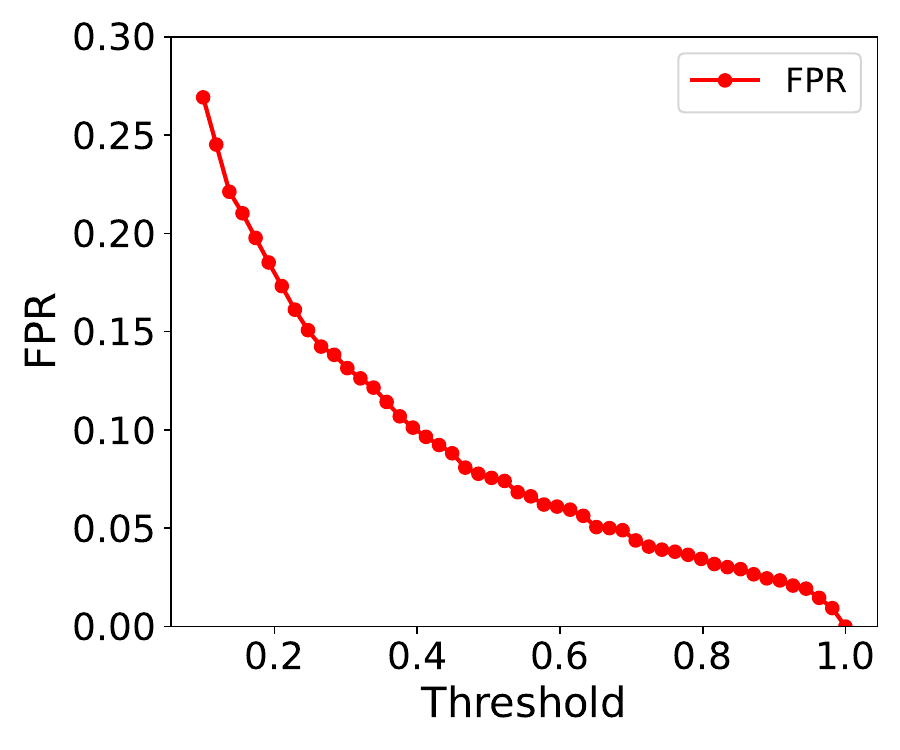}
    \caption{Variation of TPR (left) and FPR (right) as a function of threshold for $m_\Phi=15$ GeV.}
    \label{fig:tprfpr_threshold}
\end{figure}

This appendix quantifies the impact of the threshold on the TPR and FPR.  The results, in Fig.~\ref{fig:tprfpr_threshold}, are illustrated for $m_\Phi=15$ GeV. However, the understanding can be extended to other masses as well.
The TPR (left) shows an expected decrease with an increase in threshold. The interval $[0.4,0.75]$ is interesting due to the appearance of a plateau-like region in TPR and is accompanied by a similar nature, but substantially smaller values for FPR. This confirms the robustness of the choice $\tt{T}=0.5$ and any variation in this range does not substantially alter the final results.
A key point of difference between $m_\Phi=15$ GeV and the heavier masses is the relatively larger FPR as the signal is increasingly background-like.

\section{Comparison of distributions for the true positives and false negatives for $m_{\Phi}=60$ GeV}
\label{app:tprfnr60}
The appendix illustrates the true positives and false negatives for the $m_\Phi=60$ GeV benchmark.
This suggests the classifier is better at tagging events where the two b-jets are moderately separated, while events with more widely separated jets tend to get misclassified as background (false negatives).
The photon $p_T$ is generally harder than in the 15 GeV case, but high-$p_T$ photons above ~80 GeV are frequently missed. The invariant mass of the b-jet pair clusters strongly around 50–70 GeV for TPs, matching the $\Phi$ resonance. 

\begin{figure}[htb!]
    \centering
    \includegraphics[width=0.48\linewidth]{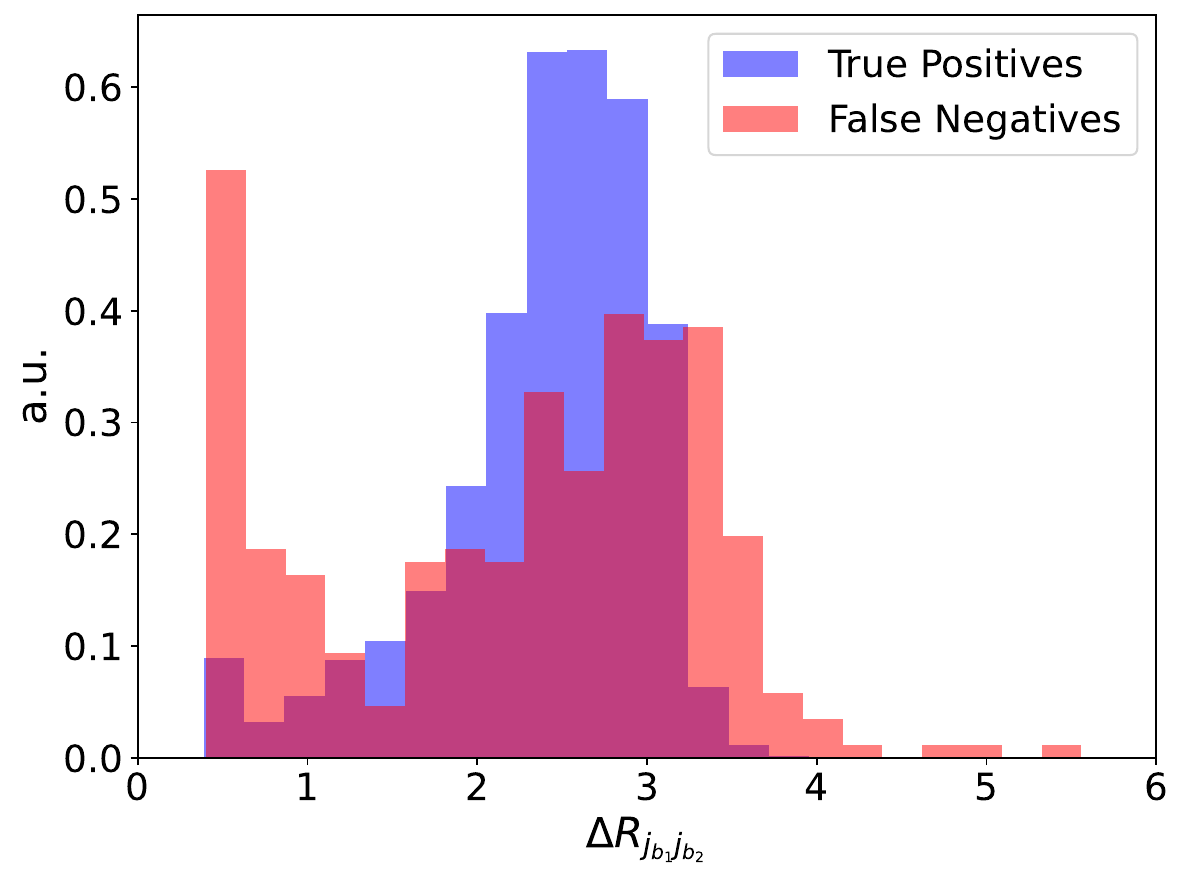}
    \includegraphics[width=0.48\linewidth]{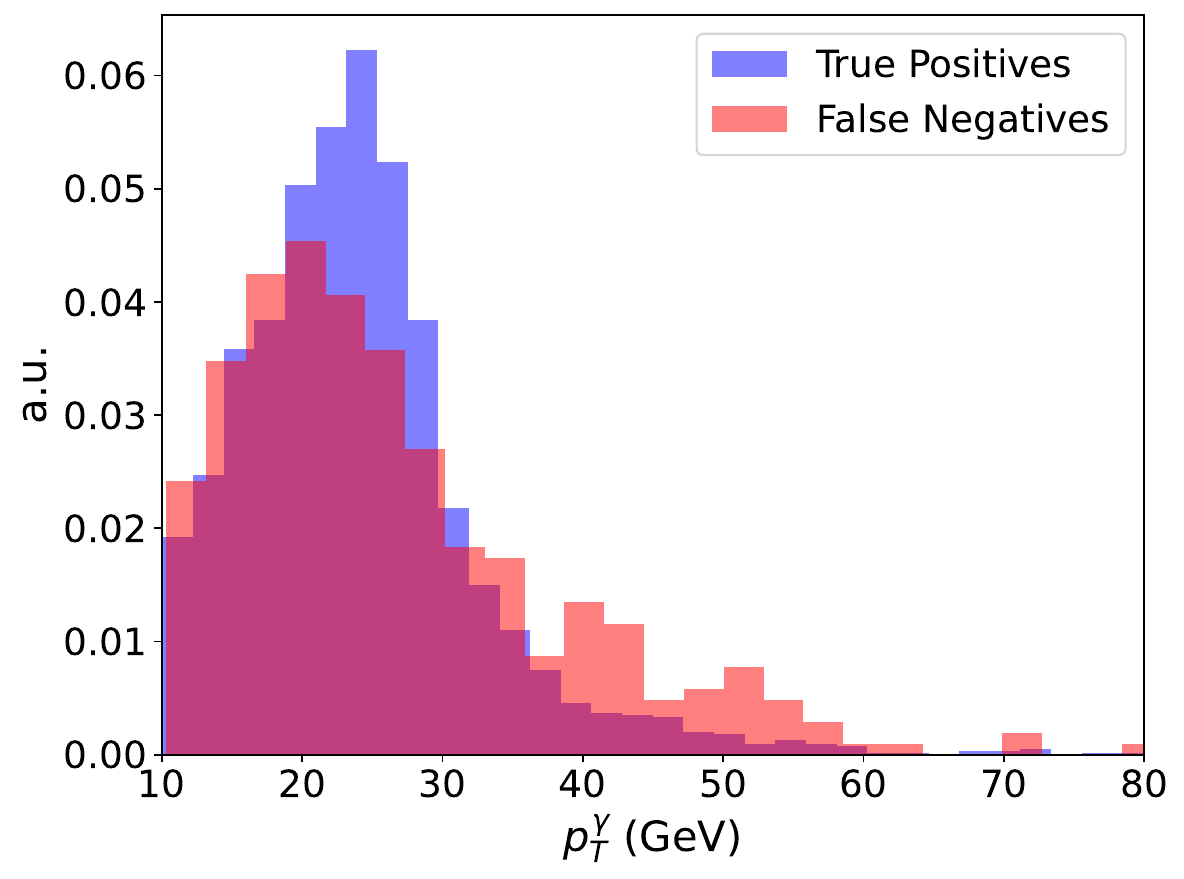} \\
    \includegraphics[width=0.48\linewidth]{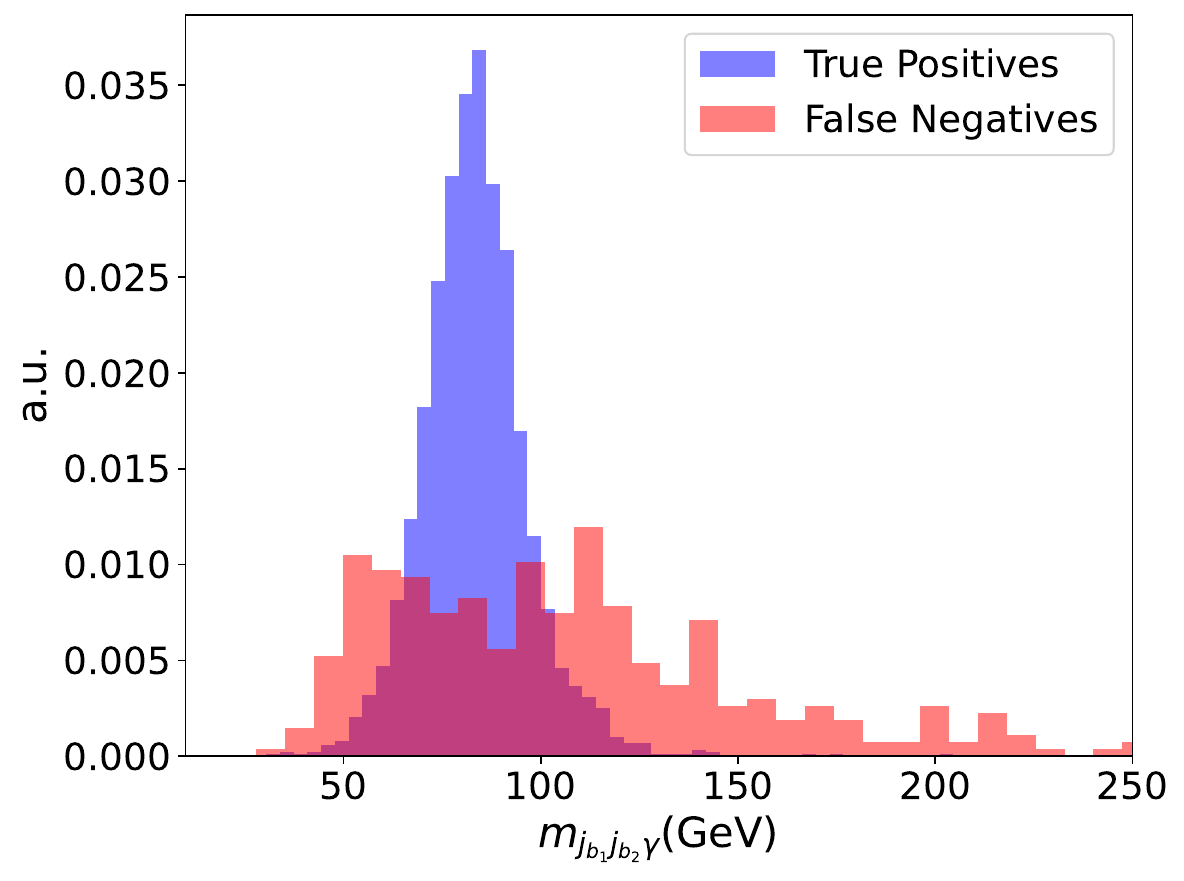}
    \includegraphics[width=0.48\linewidth]{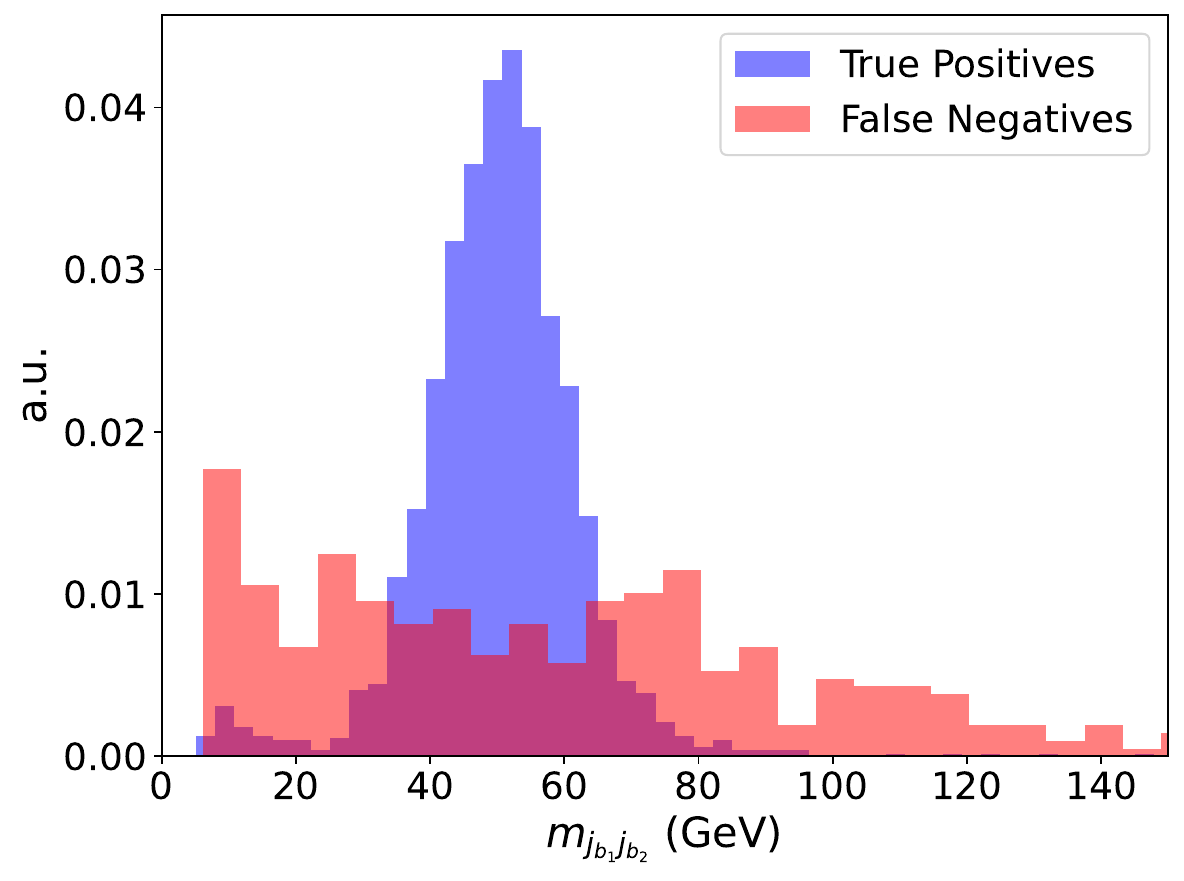} \\
    \includegraphics[width=0.48\linewidth]{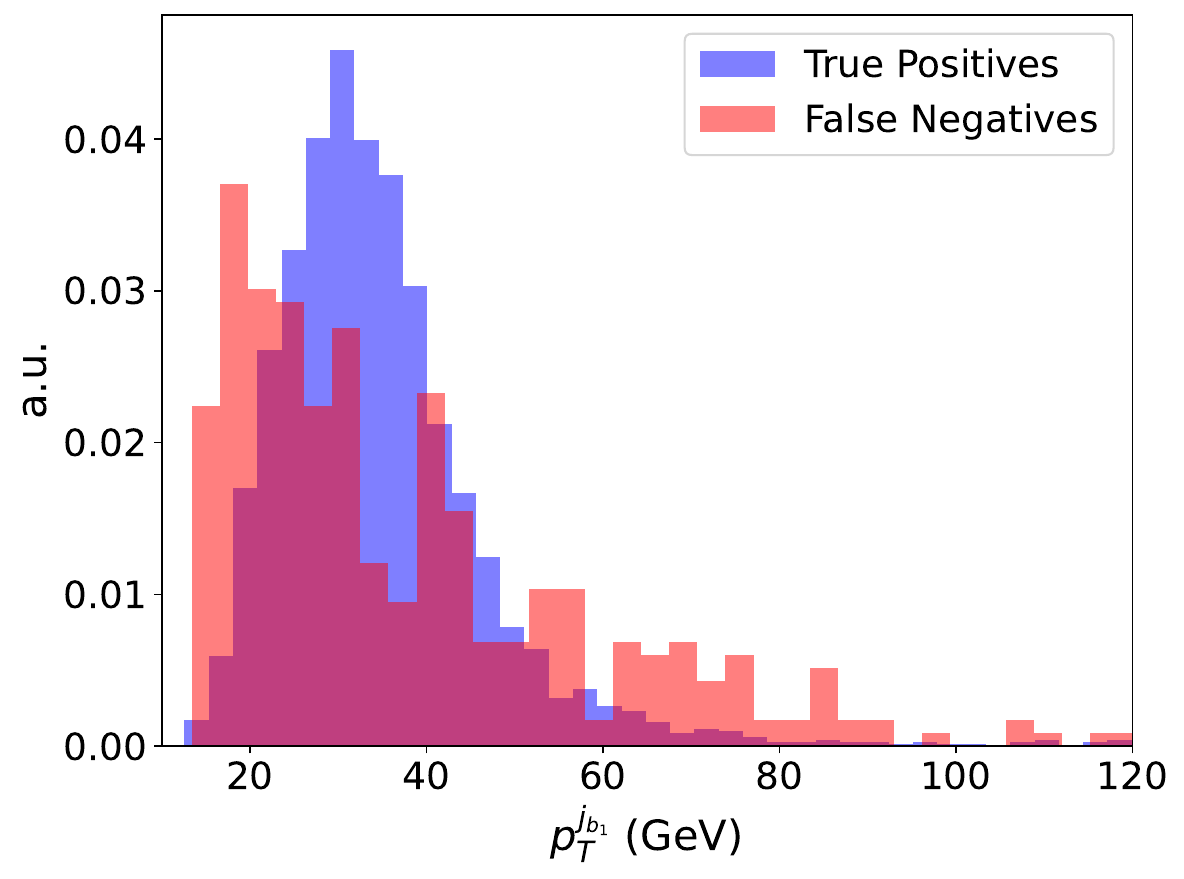}
    \includegraphics[width=0.48\linewidth]{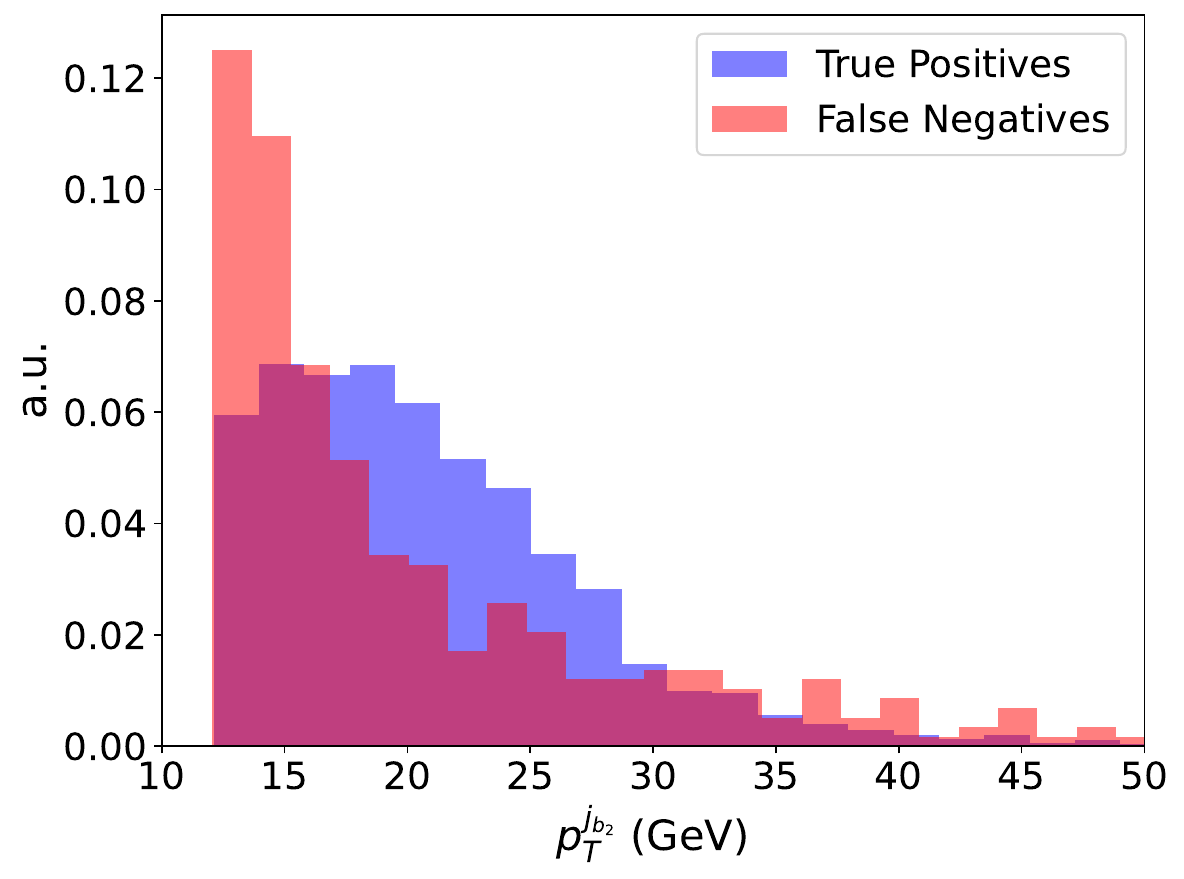}
    \caption{True Positives and False Negatives as predicted by the Gradient Boosting algorithm when $m_\Phi = 60 $ GeV}
    \label{fig:GradientBoost_60GeV}
\end{figure}

\clearpage

\end{document}